\documentclass[%
reprint,
superscriptaddress,
preprintnumbers,
nofootinbib,
nobibnotes,
 amsmath,amssymb,
 aps, prd
]{revtex4-2}

\usepackage{graphicx}
\usepackage{dcolumn}
\usepackage{bm}
\usepackage{hyperref}
\usepackage{aas_macros}
\usepackage{amsmath} 
\usepackage{physics}
\usepackage{amssymb}
\usepackage{multirow}
\usepackage{graphicx}
\usepackage[table,xcdraw]{xcolor}

\usepackage{orcidlink}

\graphicspath{{./}{figures/}}

\begin{document}

\preprint{APS/123-QED}

\title{Hyperaccreting Neutron Stars inside Massive Envelopes: The Implausibility of Thorne-\.Zytkow Objects }

\author{Patrick Chi-Kit \surname{Cheong}~\orcidlink{0000-0003-1449-3363}}
\affiliation{Unaffiliated}

\author{David \surname{Radice}~\orcidlink{0000-0001-6982-1008}}
\thanks{Alfred P.~Sloan Fellow}
\affiliation{Institute for Gravitation and the Cosmos, The Pennsylvania State University, University Park PA 16802, USA}
\affiliation{Department of Physics, The Pennsylvania State University, University Park PA 16802, USA}
\affiliation{Department of Astronomy \& Astrophysics, The Pennsylvania State University, University Park PA 16802, USA}

\author{Christopher L. \surname{Fryer}~\orcidlink{0000-0003-2624-0056}}
\affiliation{Center for Nonlinear Studies, Los Alamos National Laboratory, Los Alamos, NM 87545, USA}

\date{\today}

\begin{abstract}
The evolution of neutron stars (NSs) embedded within massive stellar envelopes is a critical phase in binary stellar evolution, potentially leading to the formation of Thorne-\.Zytkow Objects (T\.ZOs) or catastrophic collapse.
We present the first fully coupled general relativistic hydrodynamics (GRHD) simulations of hypercritical accretion onto NSs that simultaneously incorporate grey two-moment (M1) neutrino transport and an $\alpha$-chain nuclear reaction network.
By investigating four distinct progenitor evolutionary stages, we resolve the complex interplay between intense neutrino cooling, multidimensional fluid dynamics, and nuclear feedback.
Our results show that while vigorous convection is triggered in the post-shock region, the global energy budget is primarily governed by neutrino cooling, which effectively balances the accretion power.
Crucially, even though our M1 transport scheme captures neutrino absorption and localized heating, the efficient cooling sink and high ram pressure of the infalling envelope prevent the formation of any core-collapse supernova-like explosion.
We find that all nucleosynthetically processed material ($T > 5$~GK) remains strictly gravitationally bound, challenging the assumption that these systems contribute significantly to galactic nucleosynthetic yields via convective dredge-up.
The lack of sustained outflows and the persistent hypercritical accretion rates suggest that embedded NSs will rapidly exceed the Tolman-Oppenheimer-Volkoff mass limit on timescales of minutes to hours.
We conclude that these systems are not stable T\.ZOs, but are rather transient precursors to catastrophic black hole formation and potential central engines for high-energy transients.
\end{abstract}

\maketitle


\section{\label{sec:intro}Introduction}

Thorne-\.Zytkow objects (T\.ZOs) are hypothesized to form when a neutron star (NS) becomes embedded in the envelope of a massive star, typically a red giant or supergiant \cite{1975ApJ...199L..19T, 1977ApJ...212..832T}.
This embedding can occur through common-envelope (CE) evolution \cite{1978ApJ...222..269T, 2013A&ARv..21...59I, 2022MNRAS.513.4802A, 2025Ap&SS.370...11G, 2026enap....3..336O}, stellar collisions within globular clusters \cite{1992ApJ...389..546B}, or post-supernova interactions \cite{1994ApJ...423L..19L}.
In such systems, the NS accretes material from the surrounding envelope, potentially powering the system via gravitational accretion energy or exotic nuclear burning.

Following the onset of a merger, the NS spirals inward through the extended envelope of the giant or supergiant.
During this phase, frictional drag and energy deposition unbind part of the envelope while the NS approaches the stellar core.
Once the NS reaches the vicinity of the core, the accretion dynamics change qualitatively; local densities and temperatures rise sharply, and the mass flux onto the NS can become highly time-dependent and hypercritical, far exceeding the Eddington limit.
At these extreme densities, photons are essentially trapped; they provide crucial radiation pressure support but cannot transport energy outward.
Consequently, neutrino cooling emerges as the dominant mechanism regulating the accretion shock and the thermal evolution of the infalling material.
Such intense neutrino emission may even be detectable with next-generation neutrino observatories \cite{2025ApJ...984L...2M, 2025PhRvL.134r1003E}.
Understanding the physics of this phase is essential for predicting whether the system forms a T\.ZO, rapidly collapses into a black hole (BH), or triggers envelope ejection via feedback processes.

The ultimate fate of the NS-core accretion phase remains a subject of significant uncertainty, with the community broadly divided into two theoretical paradigms.
From a stellar evolution perspective, early 1D models suggested that if the NS successfully reaches the core without ejecting the envelope, the resulting T\.ZO could remain stable on secular timescales ($> 10^5$~yr) \cite{1977ApJ...212..832T}.
However, achieving this stability often required specific numerical assumptions that are difficult to reconcile with the extreme physical environment of the core.
Classic models frequently omitted or simplified neutrino microphysics \cite{1984SvA....28..187B, 1989ApJ...346..277E, 1991ApJ...380..167B, 1994ApJ...420..364B}, which are critical energy sinks during hypercritical accretion.
Without efficient neutrino cooling, the accretion energy remains trapped, artificially supporting the stellar structure against gravitational collapse.
Even in recent 1D evolutionary models, achieving stability requires specific artificial treatments of the inner boundary, such as significantly inflating the effective radius of the NS to avoid the extreme core densities where neutrino cooling would otherwise dominate \cite{2023MNRAS.524.1692F}.

Moreover, the observational necessity for long-lived T\.ZOs has been challenged.
The anomalous chemical abundances (e.g., Mo and Rb) often cited as T\.ZO signatures \cite{2018MNRAS.479.3101B, 2021A&A...646A..98T} can alternatively be produced by the $i$-process in standard stellar evolution \cite{2010ApJ...710.1557P}.
Similarly, the kinematic offsets of candidate stars are often consistent with ejections from tidally disrupted open clusters \cite{2011Sci...334.1380F}.

Conversely, from a dynamical perspective, analytic studies \cite{1993ApJ...411L..33C, 1995PhR...256...95C, 1996ApJ...459..322C, 1996ApJ...460..801F} and hydrodynamic simulations \cite{1996ApJ...460..801F, 2026ApJ...997...88A} suggest that hypercritical accretion enabled by neutrino cooling makes a long-lived T\.ZO physically untenable.
Because neutrinos efficiently evacuate the thermal energy required for hydrostatic pressure support, the accretion flow proceeds in a state of near free-fall, likely causing the NS to exceed its maximum mass and collapse into a BH on a dynamical timescale.
In this framework, the embedded NS/BH phase is a rapid, transient precursor to high-energy phenomena such as fast blue optical transients (FBOTs), ultra-long gamma-ray bursts (GRBs), and specific classes of supernovae \cite{1998ApJ...502L...9F, 2001ApJ...550..357Z, 2012ApJ...752L...2C, 2022ApJ...932...84M, 2025arXiv251009745K, 2026arXiv260323597N} (e.g. the common envelope jets supernova (CEJSN) scenario \cite{2015MNRAS.449..288P, 2019MNRAS.484.4972S, 2021MNRAS.507.1651G, 2022RAA....22e5010S}).

Recent studies have explored the dynamical stability of NS--massive star mergers \cite{2025ApJ...993...61W}, with some suggesting survival through the common envelope phase with only modest mass gain \cite{2015ApJ...798L..19M}.
Simulations have further analysed this evolution, highlighting the role of enhanced drag forces in accelerating the core merger \cite{2026arXiv260419236S}, and the impact of prescribed jets on envelope unbinding \cite{2022MNRAS.514.3212H}.
Furthermore, the high accretion rates observed in these 3D orbital studies have led to the inference that collapse due to hyperaccretion likely makes long-lived T\.ZOs physically untenable \cite{2024ApJ...977..196H, 2024ApJ...971..132E}.

However, while these global models capture the large-scale dynamics, they often lack the resolution and self-consistent microphysics required to resolve the thermodynamic structure of the final merger with the core.
Previous smaller-scale hydrodynamic simulations \cite{1996ApJ...460..801F, 2025ApJ...987...71C, 2026ApJ...997...88A} have addressed the inner engine but have largely relied on parametrised neutrino cooling and simplified NS surface treatments.
Consequently, they cannot physically capture the critical boundary layer heating at the NS surface, which could impact the highly localized convective feedback driven by nuclear burning.
To definitively determine the structural stability of the deeply embedded phase and assess the potential for nucleosynthetic dredge-up, it is imperative to employ a high-fidelity setup that incorporates realistic stellar profiles at the appropriate evolutionary stage, neutrino transport with realistic microphysics, and nuclear energy feedback within a general relativistic hydrodynamics (GRHD) framework.

The outcome of this phase likely depends on the evolutionary state of the donor star, as the stellar radius, envelope binding energy, and core structure vary significantly between hydrogen-shell burning giants and more advanced evolutionary stages.
Despite the importance of these structural factors, fully self-consistent modelling of the hyperaccreting NSs inside massive envelopes, coupling multidimensional hydrodynamics with comprehensive neutrino transport, has not been reported.

In this work, we connect the stellar structure of the donor star with the time-dependent accretion onto an embedded NS.
We investigate whether the system reaches a quasi-steady state and how this evolution depends on the stellar evolutionary stage.
Our simulations employ GRHD with grey two-moment neutrino transport and a 13-isotope $\alpha$-chain nuclear reaction network.
By utilizing the realistic equation of state, we self-consistently account for the thermodynamic contributions of radiation and ions, providing a rigorous assessment of the stability and neutrino-cooling efficiency in the deeply embedded phase.

The paper is organized as follows.
Section~\ref{sec:methods} outlines the methodology and numerical setup.
Section~\ref{sec:results} presents the simulation results, and Section~\ref{sec:discussion} summarizes our findings and discusses the broader implications.

\section{\label{sec:methods}Methods}

\subsection{\label{sec:eos}Equation of State}
We employ a hybrid equation of state constructed by combining a high-temperature nuclear EoS with a low-temperature stellar EoS as in \cite{2026ApJS..284....6C}.
At high temperatures (i.e. $T \geq 5.8~\mathrm{GK}$), we assume nuclear statistical equilibrium (NSE) and use the DD2 EoS~\cite{2010NuPhA.837..210H}, which provides thermodynamic quantities for dense, hot matter including contributions from nucleons, nuclei, and leptons.
At lower temperatures (i.e. $T < 5~\mathrm{GK}$), we adopt the \texttt{helmeos} stellar EoS~\cite{2000ApJS..126..501T}, which includes ideal gas ions, photon radiation, and degenerate or relativistic electrons and positrons, in local thermodynamic equilibrium.
In the intermediate region, $5~\mathrm{GK} \leq T < 5.8~\mathrm{GK}$, all thermodynamic quantities (pressure, internal energy, entropy, sound speed, etc.) are obtained via linear interpolation between the two EoSs.
NSE is enforced in the intermediate region.

Since the simulated domain lies deep within the stellar core, explicit radiative transfer for photons is neither necessary nor included.
The flow is highly optically thick, and we can analytically verify that the entire computational domain resides well inside the photon trapping radius, as discussed in Appendix~\ref{sec:photon_trapping}. 
The thermodynamic contribution of this trapped radiation field is naturally captured via the photon radiation pressure and internal energy components intrinsic to our hybrid equation of state.

\subsection{\label{sec:neutrinos}Neutrino Microphysics}
We employ a two-moment, energy-integrated neutrino transport scheme with explicit evolution of the neutrino number density \citep{2023ApJS..267...38C, 2024ApJ...975..116C}.
The scheme solves the zeroth and first moment equations of the neutrino radiation field, closed by the maximum-entropy closure \citep{1978JQSRT..20..541M}, and provides consistent coupling between neutrinos and the hydrodynamics through emission, absorption, and scattering source terms.
We evolve 3 species of neutrinos, namely, the electron neutrino $\nu_e$, anti-electron neutrino $\bar{\nu}_e$ and heavy-lepton neutrino $\nu_x$, where the muon and tauon neutrinos (i.e. $\nu_{\mu}, \bar{\nu}_{\mu}, \nu_{\tau}$ and $\bar{\nu}_{\tau}$) are grouped into $\nu_x$.

The neutrino microphysics is provided by \texttt{NuLib}~\citep{2015ApJS..219...24O}.
The table includes reaction rates for the charged current reactions $p + e^- \leftrightarrow n+\nu_e$ and $n+e^+ \leftrightarrow p+\bar\nu_e$;
scattering of neutrinos on protons, neutrons, $\alpha$-particles and heavy nuclei; $e^+e^- \leftrightarrow \nu\bar \nu$ and Bremsstrahlung for the heavy-lepton neutrinos only. 
The table is logarithmically spaced in neutrino energies $\epsilon$ (16 groups up to $528\,{\rm MeV}$), rest-mass density $\rho$ (82 points in $[10^6,3.2\times 10^{15}]\,{\rm g/cm^3}$) and fluid temperature $T$ (65 points in $[0.05,150]\,{\rm MeV}$), and linearly spaced in the electron fraction $Y_e$ (51 points in $[0.01,0.6]$). 
The neutrino rates are energy integrated and corrected as described in \cite{2023ApJS..267...38C}.

Neutrino transport is enabled only in regions where neutrinos are expected to interact significantly with matter.
Specifically, the neutrino-matter coupling is activated when $\rho \geq 10^5~{\rm g \cdot cm^{-3}} $ and $T \geq 0.431~{\rm MeV}$, which are typical thresholds for appreciable neutrino optical depths in dense accretion flows.
Outside this regime, neutrinos are treated as free streaming, and only their advection along characteristics is tracked without energy or momentum feedback onto the hydrodynamics.

\subsection{\label{sec:nuclear}Nuclear Network}
Fully coupled nuclear burning \citep{2026ApJS..284....6C} is employed.
We adopt the standard 13–isotope $\alpha$-chain network {\tt aprox13} \citep{1999ApJS..124..241T, 2000ApJS..129..377T}, consisting of $\rm ^4He$, $\rm ^{12}C$, $\rm ^{16}O$, $\rm ^{20}Ne$, $\rm ^{24}Mg$, $\rm ^{28}Si$, $\rm ^{32}S$, $\rm ^{36}Ar$, $\rm ^{40}Ca$, $\rm ^{44}Ti$, $\rm ^{48}Cr$, $\rm ^{52}Fe$, and $\rm ^{56}Ni$.
Free neutrons and protons are also advected as passive species but do not participate in any nuclear reactions within this network.
Nuclear reactions are activated only within the temperature range $10^6~{\rm K} \leq T \leq 5\times10^{9}~{\rm K}$.

\subsection{\label{sec:id}Initial conditions}
\subsubsection{Neutron Star Model}
The initial NS model is generated with the \texttt{RNS} code~\citep{1995ApJ...444..306S}.
Specifically, the initial configurations is constructed with central energy density $\epsilon_{c}/c^2 = 6.234\times10^{14}~\rm{g \cdot cm^{-3}}$.
The gravitational mass is $1.39985~{\rm M_{\odot}}$, where the corresponding rest masses are $1.52803~{\rm M_{\odot}}$.
This initial configuration is constructed with a fixed temperature of $0.5~\mathrm{MeV}$ and in neutrinoless $\beta$-equilibrium with DD2 EoS~\citep{2010NuPhA.837..210H}.
As now the nuclear and stellar EoSs are bridged at around $t=5~{\rm GK}$, for simplicity, we set the temperature of the NS as $0.5~{\rm MeV} \gtrsim 5~{\rm GK}$, such that the NS will be handled with the nuclear EoS.

\subsubsection{Stellar Models and Evolutionary Stages}
The stellar progenitors are generated using \texttt{MESA} (Modules for Experiments in Stellar Astrophysics, version 25.12.1)~\cite{Paxton2011, Paxton2013, Paxton2015, Paxton2018, Paxton2019, Jermyn2023}.
We evolve non-rotating stars from the pre-main sequence with an initial mass of $15\,M_\odot$ and $20\,M_\odot$ and solar metallicity ($Z = 0.014$), adopting Type II opacities appropriate for varying metal abundances.
The profiles are constructed without the presence of NS.

Nuclear burning is followed using the \texttt{approx21} reaction network, which captures the dominant energy-generating processes up to advanced burning stages while maintaining computational efficiency.
Convection is modeled using standard mixing-length theory with a mixing length parameter $\alpha_{\rm MLT} = 1.8$, adopting the Schwarzschild criterion.
Semiconvective mixing is included with an efficiency parameter $\alpha_{\rm sc} = 0.1$.

Mass loss is implemented using the Dutch wind scheme for both hot and cool phases, with a scaling factor of unity.
To ensure that the stellar structure is well resolved for subsequent hydrodynamic mapping, we employ enhanced spatial and temporal resolution controls.
Specifically, we adopt a global mesh refinement parameter $\texttt{mesh\_delta\_coeff} = 0.3$, a timestep control of $\texttt{varcontrol\_target} = 10^{-4}$, and allow up to $5\times10^4$ zones.
Additional refinement is applied in high-temperature regions ($\log T > 6$) using a reduced mesh coefficient, and adaptive mesh refinement is enforced in the core to maintain a minimum number of zones and logarithmic zoning.
Further refinement is triggered in regions with significant nuclear energy generation gradients to ensure that burning regions are adequately resolved.

\begin{figure*}
	\centering
	\includegraphics[width=\textwidth, angle=0]{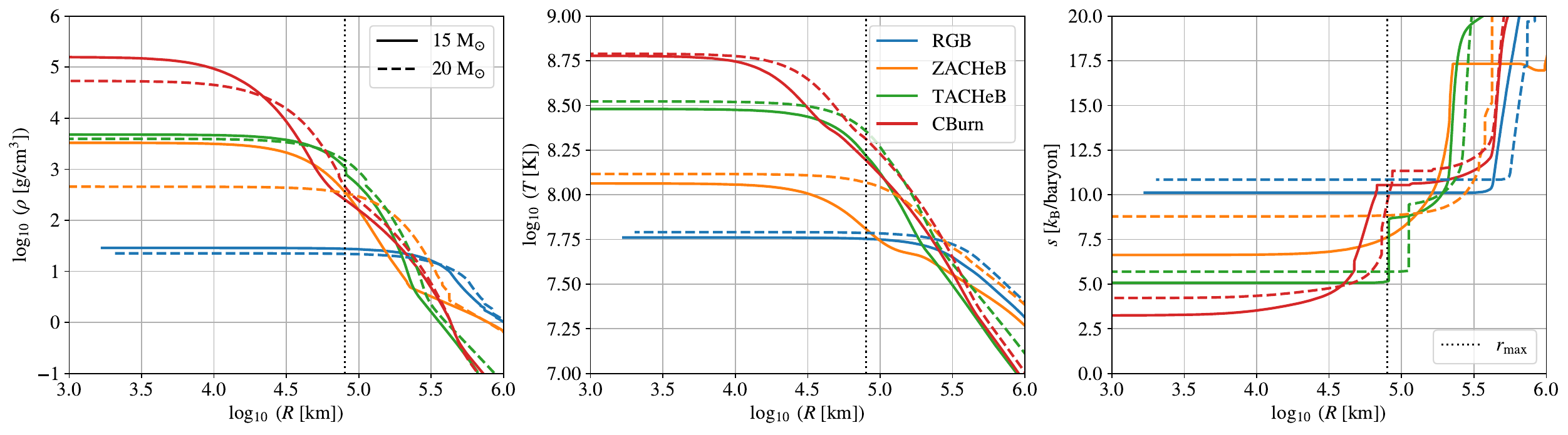}
	\caption{
		Radial profiles of density (\emph{left}), temperature (\emph{middle}), and entropy (\emph{right}) for 15 and 20 $\rm M_{\odot}$ stars at four selected evolutionary stages: RGB, ZACHeB, TACHeB, and CBurn. 
		Vertical dotted lines mark the outer edge of the computational domain used in the dynamical GRHD simulations.
		The profiles highlight the structural changes in the core across the stages, which are relevant for the dynamical simulations.
	}
	\label{fig:star_profiles}
\end{figure*}

\begin{figure*}
	\centering
	\includegraphics[width=\textwidth, angle=0]{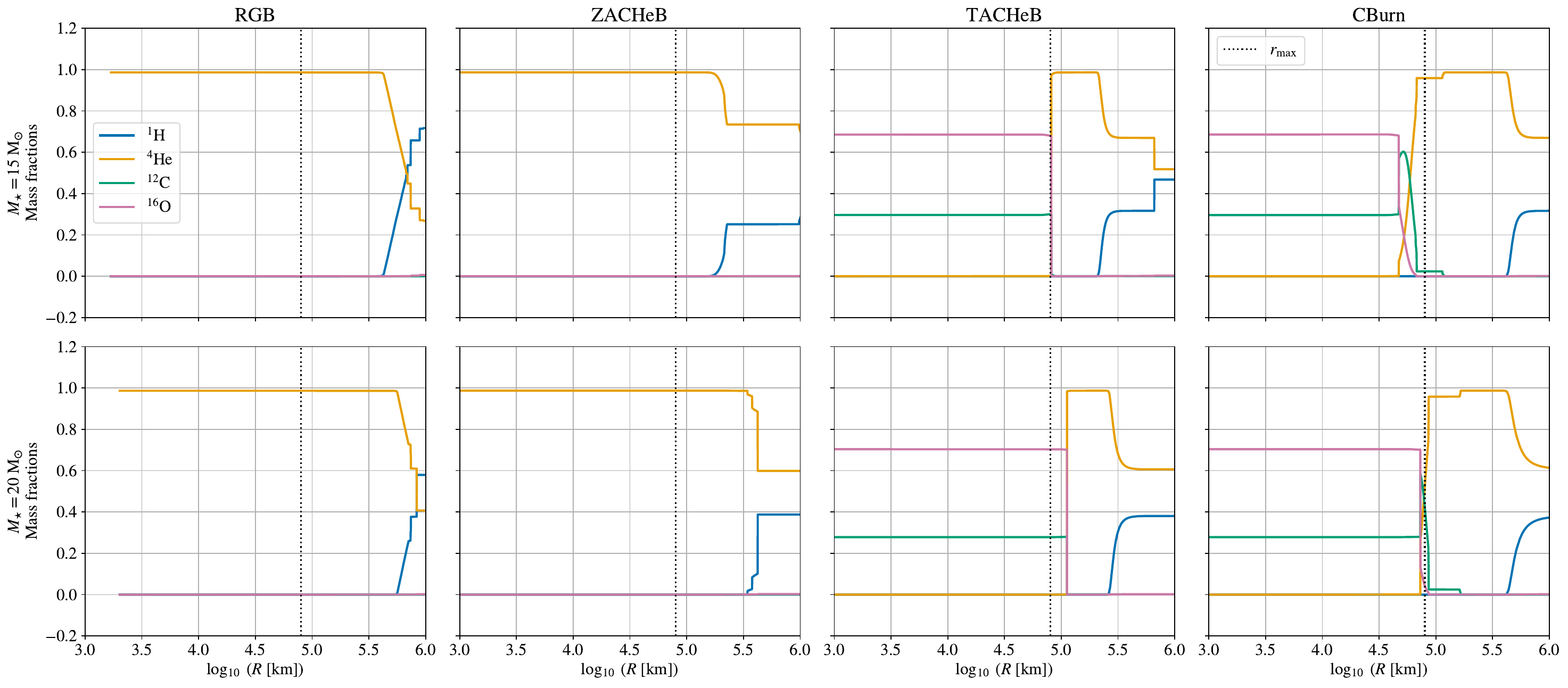}
	\caption{
		Radial composition profiles of 15 and 20 $\rm M_{\odot}$ stars at the same four evolutionary stages (RGB, ZACHeB, TACHeB, CBurn). 
		Vertical dotted lines mark the outer edge of the computational domain used in the dynamical GRHD simulations.
		Each panel shows the mass fractions of key elements (H, He, C, O), illustrating the changes in core and envelope composition relevant for nuclear burning and accretion onto the NS.
	}
	\label{fig:star_compositions}
\end{figure*}

For each mass, we extract stellar profiles at four evolutionary stages:
\begin{itemize}
	\item Red Giant Branch (RGB): corresponding to the hydrogen-shell burning phase following central hydrogen exhaustion, characterized by an inert helium core surrounded by an active H-burning shell.
	\item Zero-age core helium burning (ZACHeB): marking the onset of stable helium burning, with a well-defined helium core and moderate central density.
	\item Terminal-age core helium burning (TACHeB): corresponding to helium exhaustion, where the core is primarily composed of carbon and oxygen and could be contracted significantly.
	\item Early carbon burning (CBurn): representing the onset of advanced burning, characterized by high central density and temperature.
\end{itemize}
Figures~\ref{fig:star_profiles} and \ref{fig:star_compositions} show the density, temperature, entropy, and compositions at the core of all cases.

The four evolutionary stages considered in this work differ significantly in their physical configurations—specifically their evolutionary lifetimes, radial extents, and interior density profiles, which naturally influence the dynamics of a NS inspiral and merger.

We begin with the RGB stage, representing a post-main-sequence star with an exhausted hydrogen core and moderate envelope expansion.
While the envelope is less extended than in later stages, including RGB models provides a useful baseline for exploring lower-density, lower-accretion scenarios during an early-stage engulfment.

Following this is the ZACHeB stage, representing a stable, extended giant phase during core helium burning.
Its relatively long evolutionary duration provides a substantial time window for a binary companion to initiate a common-envelope phase and inspiral, making it a highly relevant structural configuration for modelling NS-core interactions.

As the star exhausts its core helium and progresses to the TACHeB phase, it undergoes profound structural changes.
In our stellar profiles, we find that the TACHeB radius expands to be approximately an order of magnitude larger than that of the ZACHeB stage.
This massive radial expansion naturally results in a loosely bound outer envelope.
Consequently, despite the phase having a shorter evolutionary duration than ZACHeB, the sheer physical size of the star makes it highly susceptible to Roche lobe overflow and companion engulfment.

Finally, the CBurn stage marks the onset of advanced nuclear burning.
While the star maintains a highly extended overall radius comparable to the TACHeB phase, its evolutionary lifetime is drastically shorter, making it statistically less likely for a merger to be perfectly timed to this exact epoch.
Nevertheless, as the stellar core contracts and heats up to ignite carbon, the deep interior reaches the highest densities and temperatures of the sequence.
Therefore, the CBurn model serves as a crucial physical upper bound for probing the most extreme accretion and neutrino-cooling regimes explored in this work.

Overall, this selected sequence of stellar models (RGB $\rightarrow$ ZACHeB $\rightarrow$ TACHeB $\rightarrow$ CBurn) spans a broad and physically representative parameter space.
It naturally captures the transition from moderate envelopes to highly extended giants, as well as the progression toward extreme core compactness, enabling a systematic study of hypercritical NS accretion across a full spectrum of common-envelope end states.

\subsection{\label{sec:evolution}Evolutions}
In this work, we dynamically evolve 2 stellar models with 4 different phases with the NS (total 8 models) in spherical symmetry.
In addition, we explore multidimensional effects with some models with the $1.4~{\rm M_{\odot}}$ NS and $15~{\rm M_{\odot}}$ stellar progenitor.
    We first initialize the NS on our numerical grid, and then directly map the 1D stellar profiles onto it without altering the background spacetime metric or the compact object's structure. 
    Consequently, the initial configuration is intentionally not in a hydrostatic equilibrium state; rather, the core material is immediately exposed to the deep gravitational potential of the compact object, plunging the innermost layers into a supersonic free-fall that rapidly transitions into a hypercritical accretion flow.

Our models are evolved with the \texttt{Gmunu} code, which solves GRMHD~\citep{2021MNRAS.508.2279C, 2022ApJS..261...22C} and Einstein field equations in the conformally flat approximation~\citep{2020CQGra..37n5015C}. 
Energy-integrated two-moment neutrino transport~\citep{2023ApJS..267...38C, 2024ApJ...975..116C}, and nuclear burning~\citep{2026ApJS..284....6C} modules are also activated in a fully coupled manner.
We employ the IMEXCB3a time integrator \citep{2015JCoPh.286..172C}, the Harten-Lax-van Leer (HLL) Riemann solver \citep{harten1983upstream}, and a 3rd-order reconstruction piecewise parabolic method (PPM)~\citep{1984JCoPh..54..174C}. 

In the 1D cases, the computational domain extends over $r_{\rm iso} \in [0,8 \times 10^4]~{\rm km}$ with $N_r = 128$ and up to $l_{\max}=13$ mesh-refinement levels, yielding a finest grid spacing of $\Delta r_{\rm iso} \approx 153~{\rm m}$.
The 2D simulations are performed in cylindrical coordinate $(R_{\rm iso}, z_{\rm iso})$, with a computational domain extending $R_{\rm iso}, z_{\rm iso} \in [0,8 \times 10^4]~{\rm km}$ with $N_R = N_z = 128$ and up to $l_{\max}=13$ mesh-refinement levels, which gives the finest grid spacing of $\Delta R_{\rm iso} = \Delta z_{\rm iso} \approx 153~{\rm m}$.
Note that, $r_{\rm iso}$ is the isotropic radial coordinate, while the areal circumferential radius $r_{\rm circ} = \psi^2 r_{\rm iso}$ is often used as the physical coordinate.
Below, unless explicitly stated, we refer radius as the areal circumferential radius $r_{\rm{circ}}$.

Our mesh refinement strategy utilizes a dual-layered approach combining fixed spatial nesting with dynamic feature tracking similar to \cite{2023ApJS..267...38C}. 
First, we prescribe a radially dependent window of permitted refinement levels based on spatial location. 
Specifically, for any grid block where the minimum radius satisfies $r_{\min} \leq 100~{\rm km}$, the maximum and minimum allowed refinement levels are capped at $l_{\max}$ and $l_{\max}-2$, respectively. 
More generally, for any integer $l_{\max}-2 \geq n \geq 0$, within larger radial zones where $r_{\min} \leq 2^n \times 100~{\rm km}$, the permitted refinement levels are capped at $[l_{\max}-n-2, \, l_{\max}-n]$. 
We additionally enforce the lowest baseline refinement level for blocks that intersect the outer boundaries of the computational domain. 
Second, within these radially allowed bounds, the grid adaptively adjusts to physical features by tracking sharp gradients and jumps in pressure, specific entropy, and radial velocity.
This dynamic layer ensures optimal resolution along both the NS surface and the evolving shock front.

While excising the NS is possible and effective for studies focused primarily on shock and outer-flow dynamics, the construction of physically consistent inner boundary conditions is non-trivial.
Inappropriate boundary conditions intended to mimic the NS surface can introduce spurious momentum injection (e.g. \cite{2002ApJS..143..539Z}).
More sophisticated approaches either require detailed knowledge of the NS’s hydrostatic structure \cite{2026ApJ...997...88A} or avoid explicit modelling of the surface altogether \cite{2025ApJ...985...62P}.
To eliminate these ambiguities, we allow the NS surface to evolve freely, but freeze the hydrodynamical and radiation variables in the NS interior where $r<8~{\rm km}$ during the evolution. 
This treatment is similar to \cite{2024ApJ...960..116A}, where they fix only the temperature.

\section{\label{sec:results}Results}

\subsection{\label{sec:1d_results}One-dimensional simulations}

\begin{figure*}
	\centering
	\includegraphics[width=\textwidth, angle=0]{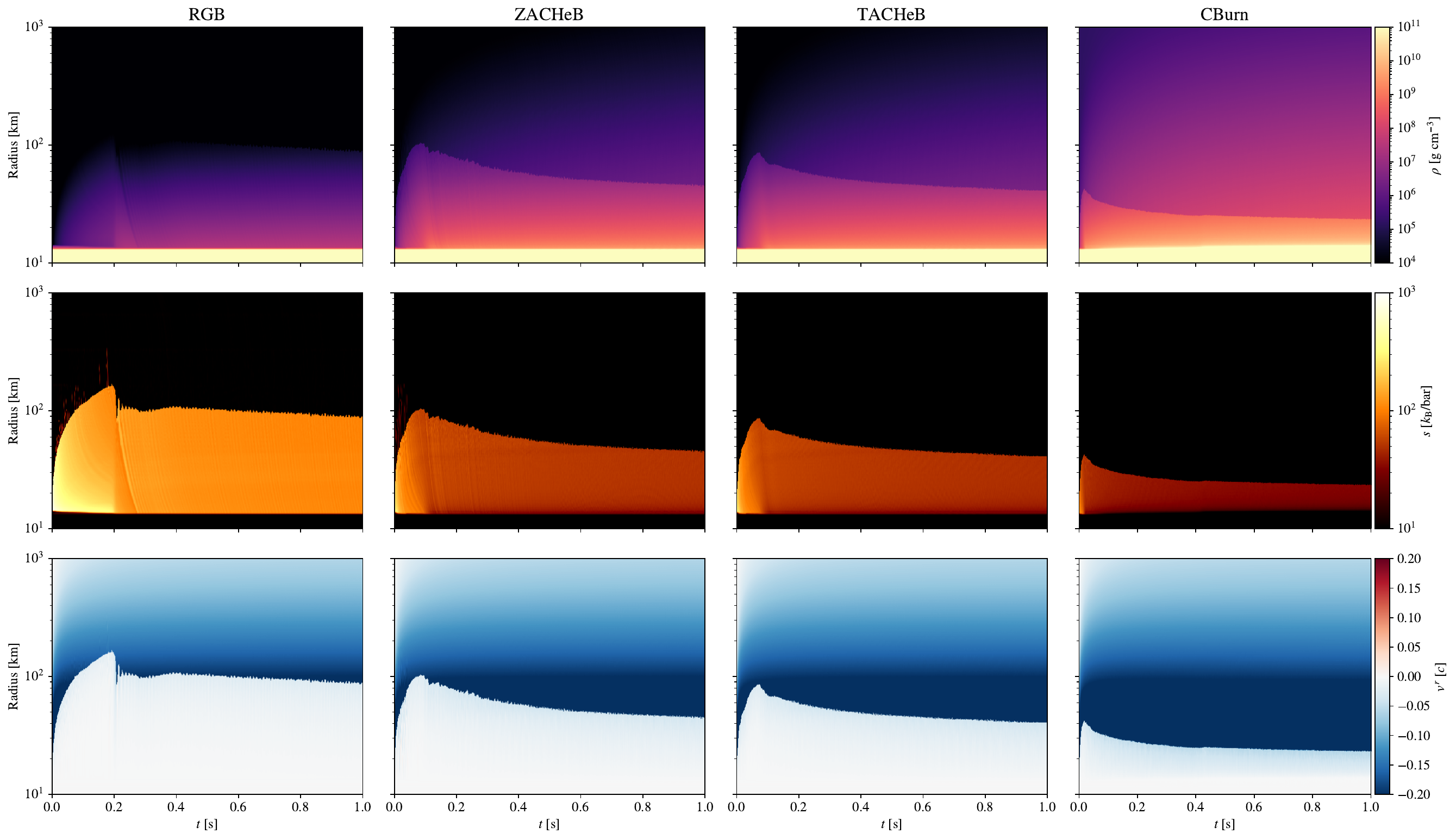}
	\caption{
		Profiles of rest-mass density $\rho$, entropy $s$, and radial velocity $v^{r}$ (\emph{top to bottom}) for models with $M_{\rm NS} = 1.4~{\rm M_{\odot}}$ and $M_{\star} = 15~{\rm M_{\odot}}$ in all four stages (\emph{left to right}). 
		In all cases, an accretion shock forms, indicated by the minimum in $v^{r}$. 
	}
	\label{fig:dynamics}
\end{figure*}

Figure~\ref{fig:dynamics} presents the time evolution of the $M_{\rm NS} = 1.4~{\rm M_{\odot}}$ and $M_{\star} = 15~{\rm M_{\odot}}$ models across all four evolutionary stages.
We show profiles of rest-mass density $\rho$, specific entropy $s$, and radial velocity $v^{r}$.
At early times, the infalling stellar material impacts the NS surface and accumulates, forming an accretion shock identified by the minimum in $v^{r}$.
The post-shock region initially exhibits a negative entropy gradient, which gradually flattens and becomes weakly positive at later times.
The duration of this phase depends on the progenitor structure, with higher core density models transitioning more rapidly.

\begin{figure*}
	\centering
	\includegraphics[width=\textwidth, angle=0]{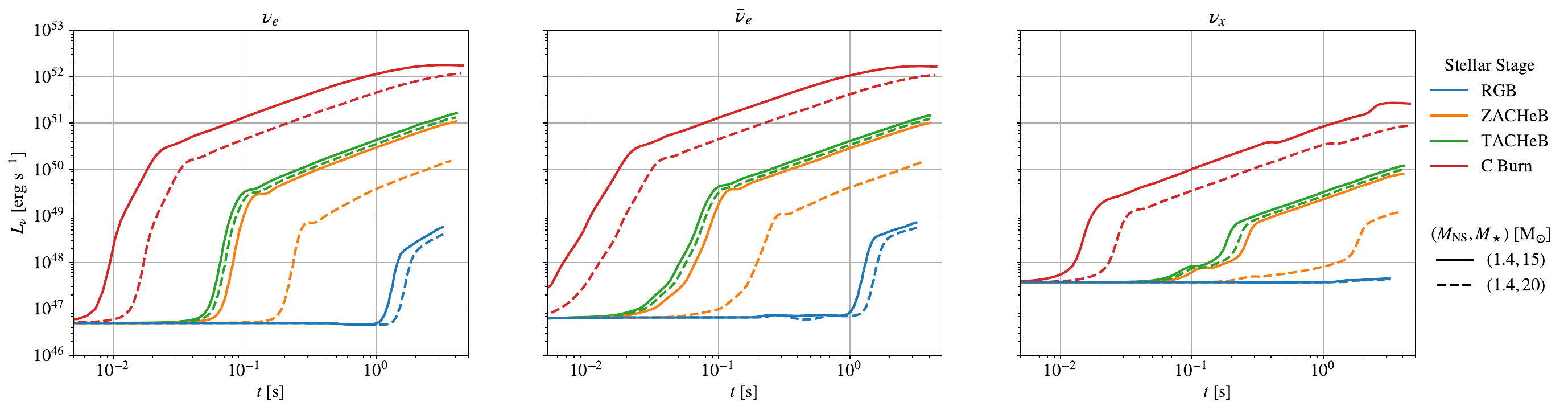}
	\caption{
		Time evolution of neutrino luminosities for $\nu_e$, $\bar{\nu}_e$, and $\nu_x$ (\emph{left to right}) extracted at $r=500~{\rm km}$ across all models.
	}
	\label{fig:nu_lum}
\end{figure*}

\begin{figure*}
	\centering
	\includegraphics[width=\textwidth, angle=0]{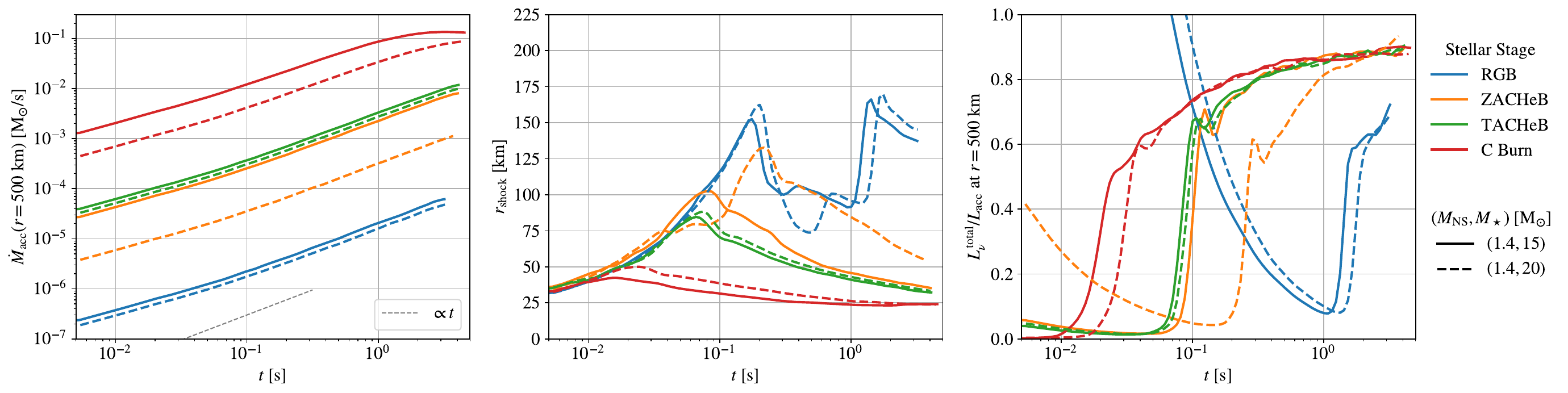}
	\caption{
		Time evolution of mass accretion rate (\emph{left}), shock radius (\emph{middle}), and the ratio of total neutrino luminosity to accretion power (\emph{right}) across all 1D models.
	}
	\label{fig:shock_Mdot_Lratio}
\end{figure*}

The neutrino luminosities are shown in Figure~\ref{fig:nu_lum}.
At early times, all neutrino species exhibit a rapid rise as infalling material compresses and heats near the NS surface.
As the shock approaches a quasi-stationary configuration, the growth rate of the luminosity decreases, although $L_{\nu}$ continues to increase due to the rising accretion rate (Figure~\ref{fig:shock_Mdot_Lratio}).
Higher core density models exhibit a more rapid rise in luminosity.
At late times, all models show the hierarchy $L_{\nu_{\rm e}} \approx L_{\bar{\nu}_{\rm e}} > L_{\nu_x}$.
The baseline emission (before the comparable rise in luminosity) is likely due to the initial heating of the NS surface, which is very common in GRHD simulations~\cite{2025PhRvD.111b3049G} when the NS surface is fully resolved.
This initial burst is a numerical relaxation of the mapped stellar material onto the NS surface and does not impact the late-time quasi-steady accretion results.
As time progresses, when the system reach quasi-steady state, the massive influx of infalling matter drives a hypercritical accretion flow that powers an overwhelming neutrino luminosity, completely overriding the initial relaxation signature (see Figure~\ref{fig:nu_lum}). 
Because the long-term state of the system is dynamically governed by the continuous, high-ram-pressure mass supply from the outer stellar envelope, the flow loses memory of its initial transient setup and settles into a robust, self-consistent profile.
It is worth pointing out that, while the neutrino emission is very strong here, the NS is relatively cold compared to the proto-NS formed in the standard core-collapse supernovae, and most of the neutrino emission is powered by accretion, as we will discuss below.

The mass accretion rate at $r=500~{\rm km}$ is primarily determined by the progenitor core density.
Lower-density models exhibit systematically lower $\dot{M}_{\rm acc}$ (Figure~\ref{fig:shock_Mdot_Lratio}).
Over the duration of the simulations, $\dot{M}_{\rm acc}$ increases approximately linearly with time.
This behaviour reflects the time-dependent infall of stellar material: for an initially nearly uniform density profile, $M_{\rm enc} \propto r^3$ and the free-fall time scales as $t \propto r^{3/2}$, leading to $\dot{M} \propto t$.
Nevertheless, the accretion rates saturated at $\gtrsim 10^{-1}~{\rm M_{\odot}/s}$ at $t \sim 3~{\rm s }$ for the strongest accreting cases.

We define the mass accretion rate as
\begin{equation}
	\dot{M}_{\rm acc} = - \int_{S_r} \rho u^r \sqrt{-g} \, \dd \Omega,
\end{equation}
where $u^{\mu}$ is the fluid four-velocity and $g$ is the determinant of the spacetime metric. 
This is evaluated at $r=500~{\rm km}$.

The shock radius, defined as the location of the minimum radial velocity, is shown in Figure~\ref{fig:shock_Mdot_Lratio}.
At late times, the shock radius correlates with the accretion rate and therefore with the progenitor structure.

To quantify the energetics, we compare the total neutrino luminosity to the accretion power.
The accretion power is estimated as
\begin{equation}
	L_{\rm acc}(r) \approx \left[ \alpha(r) - \alpha(r_{\rm NS}) \right] \dot{M}_{\rm acc}(r)\, c^2,
\end{equation}
where $\alpha(r) - \alpha(r_{\rm NS})$ represents the specific binding energy released between radius $r$ and the NS surface. 
In the Newtonian limit, this reduces to $\sim GM_{\rm NS}/R_{\rm NS}$. 
The ratio $L_{\nu}^{\rm total} / L_{\rm acc}$ therefore measures the efficiency with which accretion power is radiated via neutrinos.

For the ZACHeB, TACHeB, and CBurn models, neutrino cooling is efficient, with $\gtrsim 85\%$ of the accretion power radiated in neutrinos.
In contrast, the RGB models do not reach a well-defined quasi-steady state within the simulated time.
At early times ($t \lesssim 0.1~{\rm s}$), the RGB model exhibits an efficiency value exceeding unity.
This feature is a direct consequence of the initial numerical transient discussed above; before the steady-state accretion flow is fully established, the neutrino emission is dominated by the ``baseline'' thermal relaxation of the newly resolved NS surface, meaning the neutrino luminosity is not yet strictly tied to the instantaneous accretion power.
Furthermore, due to its significantly lower mass accretion rate, the RGB model possesses a much longer hydrodynamical timescale to completely flush out this initial transient and fill the post-shock volume compared to the more compact, high-$\dot{M}$ progenitors. 
While the simulation duration captures the clear downward trajectory of this transient toward a sub-unity equilibrium, the cooling efficiency for the RGB case remains in a late-stage relaxation phase by the end of the run.

\begin{figure*}
	\centering
	\includegraphics[width=\textwidth, angle=0]{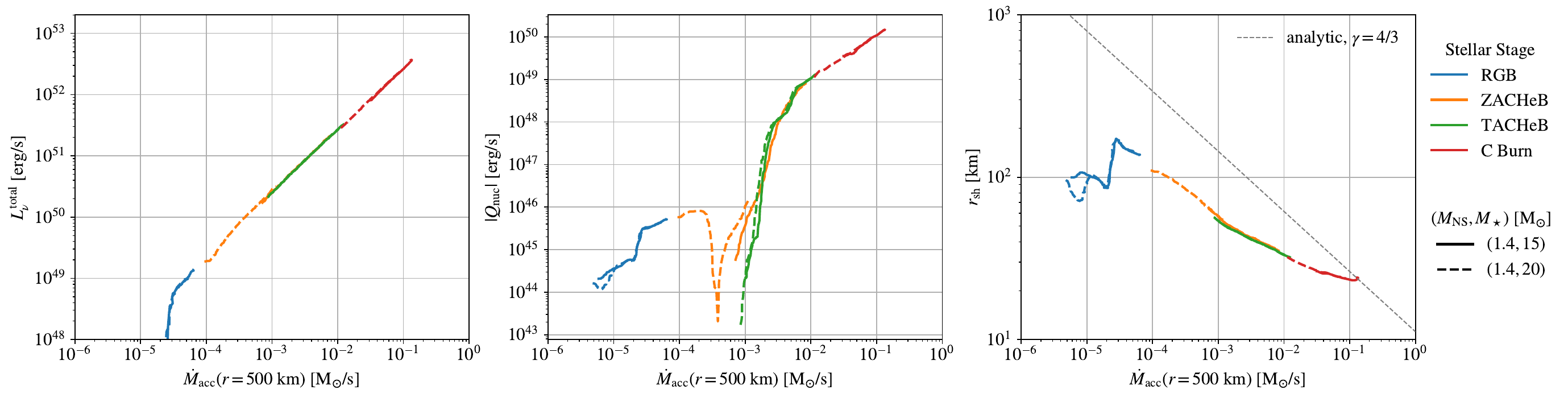}
	\caption{
		Total neutrino luminosity (\emph{left}), total nuclear energy generation rate (\emph{middle}), and shock radius (\emph{right}) as functions of the mass accretion rate.
		Only late-time evolution ($t > 0.5~{\rm s}$) is shown.
		Shock radius estimated analytically in \cite{1989ApJ...346..847C, 1996ApJ...460..801F} is shown in the grey line.
	}
	\label{fig:eng_rsh_vs_mdot}
\end{figure*}

Figure~\ref{fig:eng_rsh_vs_mdot} shows the dependence of key quantities on the accretion rate.
For the ZACHeB, TACHeB, and CBurn models, the total neutrino luminosity scales with $\dot{M}_{\rm acc}$, while the RGB models show little variation.
The total nuclear energy generation rate where $T<5~{\rm GK}$ increases with accretion rate but remains orders of magnitude smaller than the neutrino luminosity in all cases, which is consistent with \cite{2026ApJ...997...88A}.
Note that NSE is assumed when $T \geq 5~{\rm GK}$, we therefore count the nuclear energy generation rates only below that temperature.
This temperature cutoff physically accounts for the localized ``dip'' observed in the nuclear heating rate ($Q_{\rm nuc}$) around $\dot{M}_{\rm acc} \sim 10^{-3}~M_{\odot}/{\rm s}$ (Figure~\ref{fig:eng_rsh_vs_mdot}, middle panel).
At these specific intermediate accretion rates, the compressed post-shock region, which typically dominates the total nuclear heating, reaches temperatures exceeding $5~{\rm GK}$.
Consequently, this material enters the NSE regime and its net energy generation rate is no longer counted in our sub-NSE budget (as illustrated by the dashed lines in Figure~\ref{fig:compare_2d_eng_vs_t}).
As the accretion rate increases further ($\dot{M}_{\rm acc} \gtrsim 10^{-3}~M_{\odot}/{\rm s}$), intense compressional heating in the infalling \emph{pre-shock} layers elevates their temperatures to the vigorous burning regime just below $5~{\rm GK}$, allowing pre-shock nuclear heating to overtake the budget and drive the subsequent recovery of $Q_{\rm nuc}$.
The shock radius also exhibits a qualitative correlation with $\dot{M}_{\rm acc}$.

Across all models, the accretion rate—and therefore the overall flow structure—is primarily set by the progenitor core density, which acts as the main control parameter of the system.

We compare our quasi-steady shock radii to the analytic predictions of \cite{1989ApJ...346..847C, 1996ApJ...460..801F}, as shown in the right panel of Figure~\ref{fig:eng_rsh_vs_mdot}.
We find a clear mass-accretion dependence: at the highest accretion rates, our numerically determined shock radii essentially overlap with the analytic solutions, whereas at lower accretion rates, our shock radii are systematically smaller.

This convergence at high $\dot{M}_{\rm acc}$ is a physically expected outcome of the thermodynamics.
The analytic framework \cite{1989ApJ...346..847C} assumes an idealized, purely radiation-dominated fluid ($\gamma = 4/3$).
In our most hypercritical models, the extreme post-shock temperatures ensure that photon radiation pressure dominates the thermodynamic budget.
In this limit, the complex microphysics of our hybrid EoS become secondary, and the flow asymptotically approaches the idealised analytic limit.

Conversely, at lower accretion rates, the post-shock temperatures are reduced, and the flow is not strictly photon radiation dominated.
In this transitional regime, the physical complexities of our model drive systematic deviations from the analytic approximations, resulting in a more compact post-shock envelope.
First, the microphysical components of our hybrid EoS, such as nuclear dissociation and electron-positron pair creation, absorb a non-negligible fraction of the internal energy, effectively softening the equation of state relative to an idealized $\gamma = 4/3$ gas.
Second, whereas the analytic framework relies heavily on simplified power-law neutrino emissivities for thermal cooling, our two-moment neutrino transport solver naturally better captures the dominant cooling channels in the dense inner engine.
Third, because our simulations employ GRHD, the stronger gravitational potential near the NS requires a steeper pressure gradient to support the settling envelope compared to Newtonian gravity, naturally pulling the accretion shock deeper into the gravitational well.
Together, these microphysical and relativistic effects combine to substantially reduce the steady-state shock radius relative to the idealized analytic predictions.

\begin{figure}
    \centering
    \includegraphics[width=\columnwidth, angle=0]{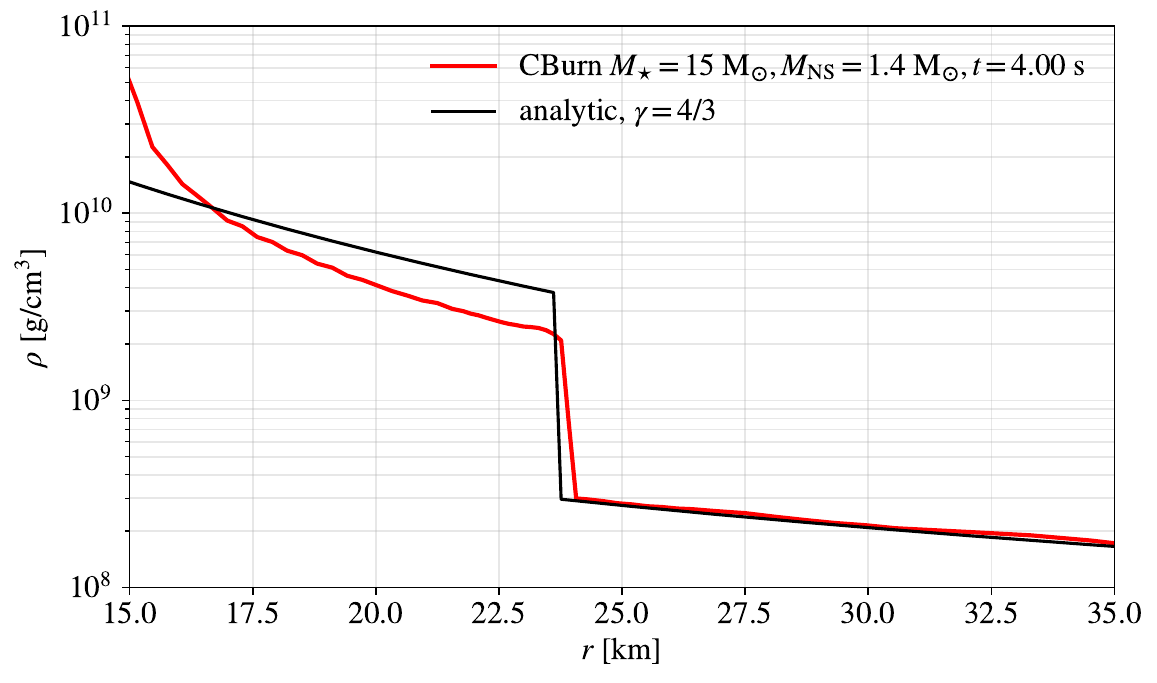}
    \caption{
        Comparison of the density profile of the CBurn model with $M_{\star}=15~{\rm M_{\odot}}, M_{\rm NS}=1.4~{\rm M_{\odot}}$ at $t=4~{\rm s}$ (red line) and the analytic model constructed in \cite{1989ApJ...346..847C, 1996ApJ...460..801F}.
    }
    \label{fig:1d_rho_profile}
\end{figure}

To further validate the thermodynamic structure of our accretion flow, we select a late-time snapshot ($t=4$~s) from our most highly accreting model (the CBurn model with $M_{\star}=15~{\rm M_{\odot}}, M_{\rm NS}=1.4~{\rm M_{\odot}}$).
In this extreme limit, the flow is radiation-dominated, providing an ideal regime to directly test our numerical framework against the classic analytic solutions derived by \cite{1989ApJ...346..847C, 1996ApJ...460..801F}.
Figure~\ref{fig:1d_rho_profile} compares the radial density profile of this model against the exact analytic curve.
The profiles show excellent agreement in the unshocked region ($r \gtrsim 24$~km), where the material is essentially in free-fall.
In the post-shock region, the profiles agree qualitatively, capturing the same macroscopic shock radius and structural trends.
The steep density enhancement deep in the interior ($r \lesssim 17.5$~km) further highlights our model's advanced microphysics and resolving the NS surface, where realistic neutrino cooling removes pressure support and drives stronger local compression than the idealized analytic approximations.

\subsection{\label{sec:2d_results}Two-dimensional simulations}

The one-dimensional models exhibit a negative entropy gradient in the post-shock region at early times, indicating conditions that may lead to multidimensional instabilities.
To investigate these effects, we perform two-dimensional simulations for the $M_{\rm NS} = 1.4~{\rm M_{\odot}}$ and $M_{\star} = 15~{\rm M_{\odot}}$ models.

\begin{figure*}
	\centering
	\includegraphics[width=\textwidth, angle=0]{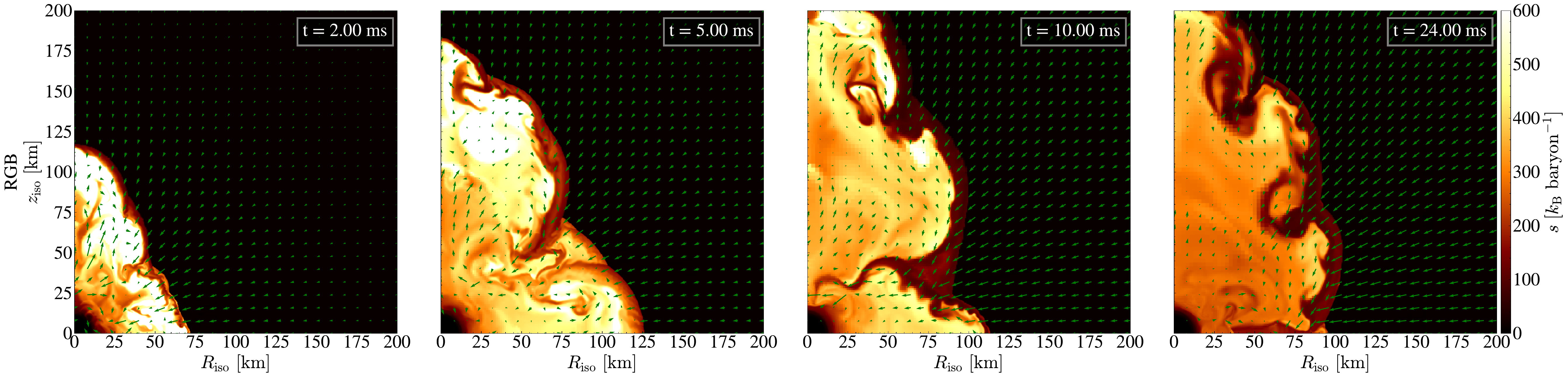}
	\includegraphics[width=\textwidth, angle=0]{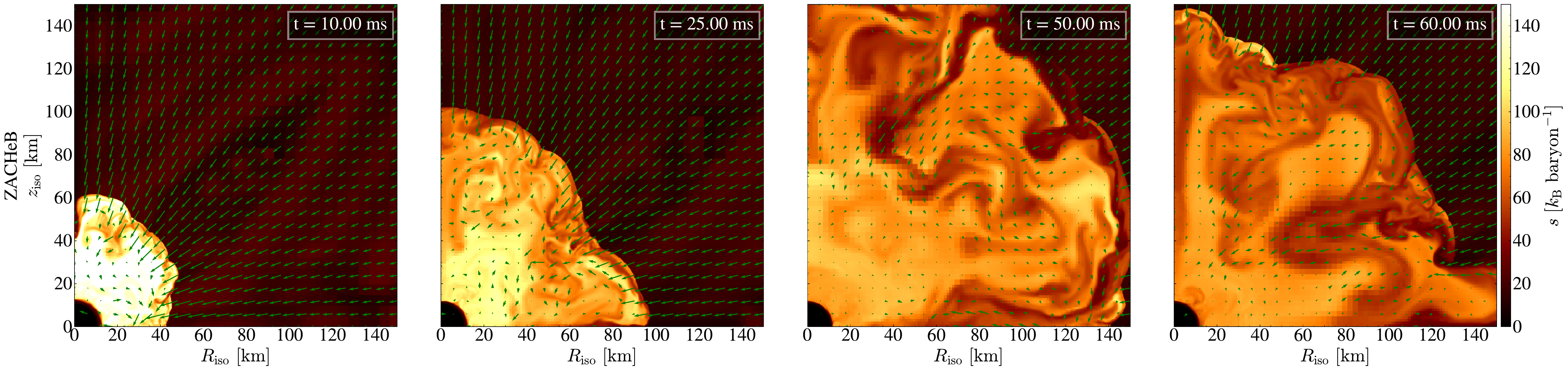}
	\includegraphics[width=\textwidth, angle=0]{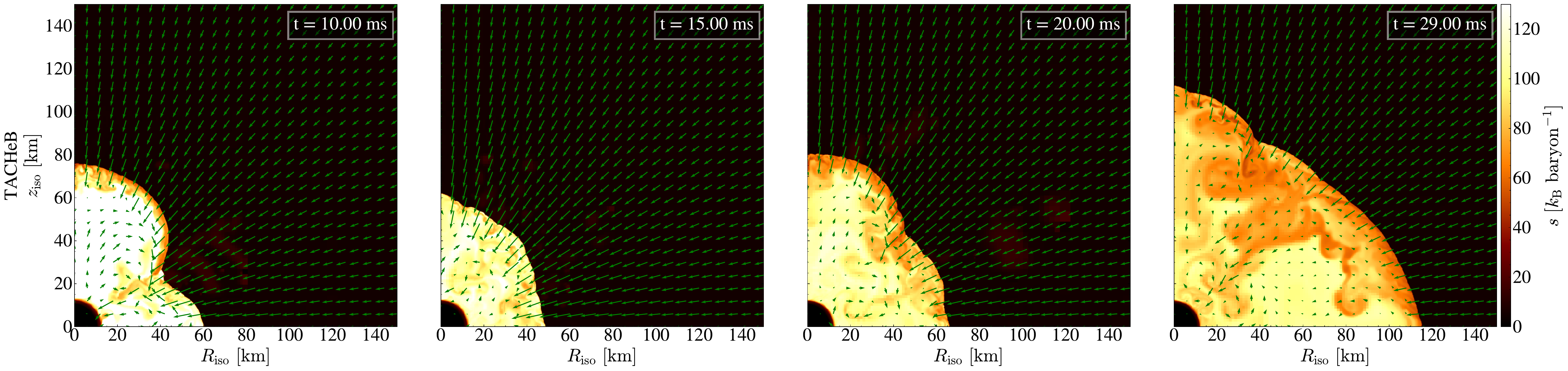}
	\includegraphics[width=\textwidth, angle=0]{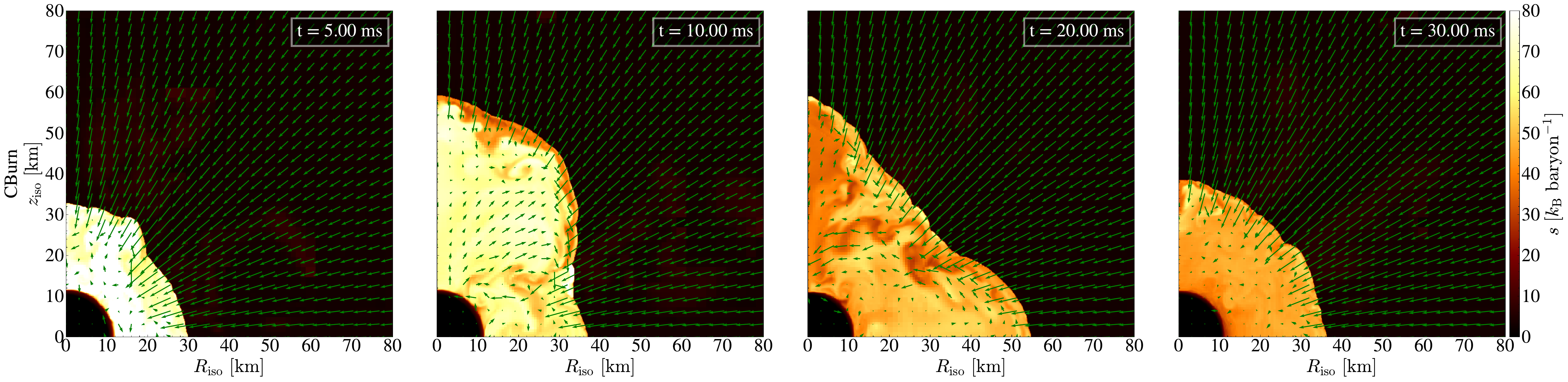}
	\caption{
		Specific entropy ($s$) profiles for different models at various times.
		Green arrows indicate the velocity vectors.
		Domain size and colour scales vary between rows for clarity.
	}
	\label{fig:2d_ent_profiles}
\end{figure*}

Figure~\ref{fig:2d_ent_profiles} shows snapshots of the entropy distribution and velocity field.
In all cases, deviations from spherical symmetry develop shortly after shock formation.
The degree and persistence of asymmetry depend on the progenitor structure.
In the CBurn model, although asymmetries appear at early times, the flow becomes nearly spherical by $t \gtrsim 30~{\rm ms}$ and remains stable thereafter.
In contrast, lower accretion rate models exhibit more prolonged non-spherical structures.

\begin{figure*}
    \centering
    \includegraphics[width=\textwidth, angle=0]{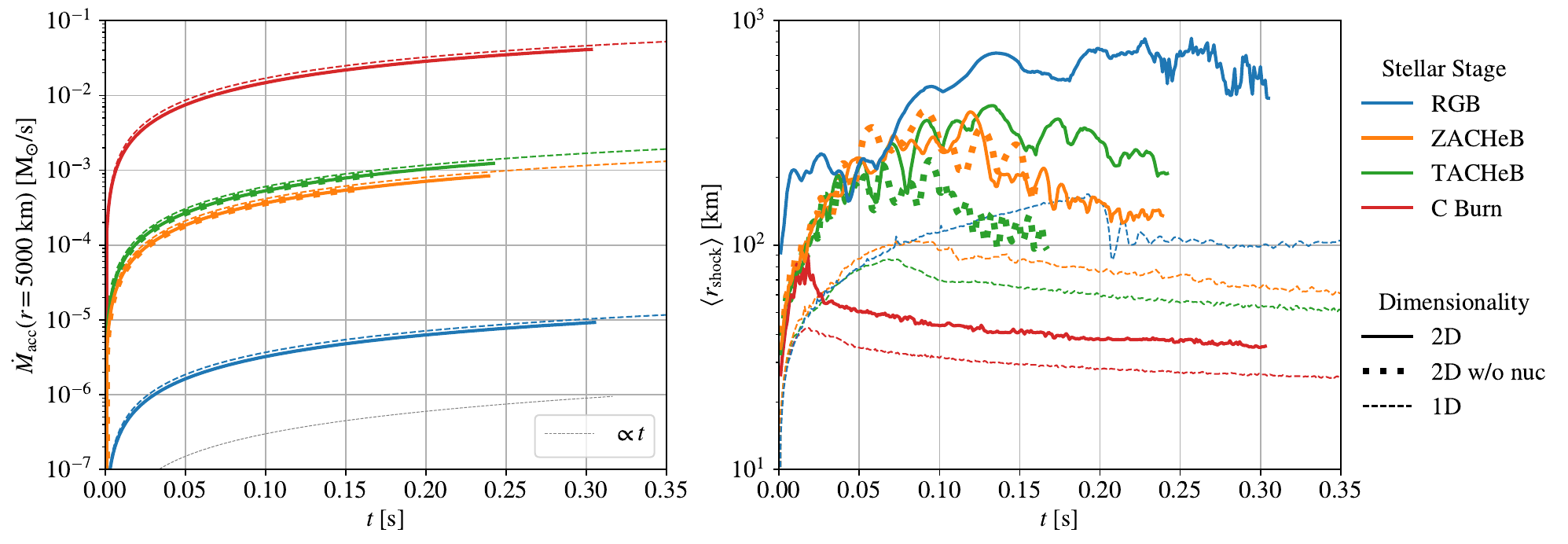}
    \caption{
        Time evolution of the mass accretion rate (\emph{left}) and the angle-averaged shock radius (\emph{right}) for the 2D (solid lines) and 1D (dashed lines) models. 
        The 2D models for ZACHeB and TACHeB without nuclear reactions are also shown as dotted lines.
        While the dotted lines in the left panel overlap completely with the solid lines, indicating that nuclear burning does not alter the bulk mass accretion rate, the right panel demonstrates that nuclear feedback in the TACHeB model significantly expands the angle-averaged shock radius compared to its non-nuclear counterpart.
    }
    \label{fig:compare_2d_shock_Mdot}
\end{figure*}

We find similar mass accretion rates measured at $5000~{\rm km}$ in both the 1D and 2D cases (see Figure~\ref{fig:compare_2d_shock_Mdot}). 
We evaluate the accretion rate at $5000~{\rm km}$ here, rather than at $500~{\rm km}$ as done previously, because the vigorous convective plumes in the 2D RGB model extend across the $500~{\rm km}$ region.
Evaluating the flux further out avoids the turbulent fluctuations associated with these convective up- and down-flows.
Because the system reaches a quasi-steady state, the bulk accretion rates at these two radii are essentially identical, with the exception of the 2D RGB case.
Furthermore, our 2D simulations without nuclear reactions (dotted lines) demonstrate that nuclear burning does not noticeably alter the bulk mass accretion rate, as these lines essentially overlap with the fully coupled models.

However, the spatial extent of the shock is profoundly altered by both multidimensional effects and nuclear feedback. 
Despite the similar accretion rates, the angle-averaged shock radius is located systematically further out in the 2D models compared to their 1D counterparts, as shown in Figure~\ref{fig:compare_2d_shock_Mdot}.
The angle-averaged shock radius is defined as
\begin{equation}
    \langle r_{\rm shock} \rangle = \frac{\int_0^{\pi/2} r_{\rm shock}(\theta) \sin\theta \, \dd \theta}{\int_0^{\pi/2} \sin\theta \, \dd \theta},
\end{equation}
where $r_{\rm shock}(\theta)$ is the radius at which the specific entropy undergoes a sharp jump along a given polar angle $\theta$.

Crucially, by comparing the fully coupled models to the runs without nuclear networks, we find that the dynamical impact of nuclear burning depends heavily on the evolutionary stage. 
In the ZACHeB model, nuclear reactions do not significantly alter the global shock dynamics, as the shock radii for the fully coupled and non-nuclear runs remain comparable. 
In contrast, the carbon/oxygen burning in the post-shock region of the 2D TACHeB model provides substantial thermal pressure and convective feedback, pushing the angle-averaged shock radius significantly further outward compared to the identical setup without nuclear burning. 
This demonstrates that while neutrino cooling ultimately regulates the mass accretion rate, extreme nuclear feedback in advanced evolutionary cores can strongly modulate the multidimensional shock morphology.

\begin{figure*}
	\centering
	\includegraphics[width=\textwidth, angle=0]{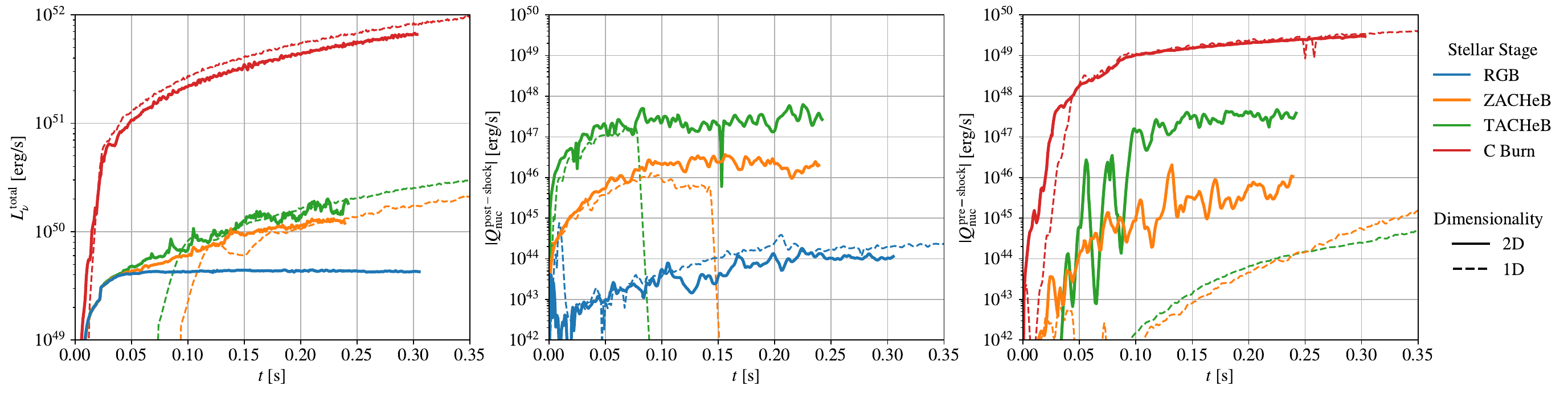}
	\caption{
		Time evolution of the total neutrino luminosity (\emph{left}) and the total energy generation rate due to nuclear reactions in the post-shock (\emph{middle}) and pre-shock (\emph{right}) regions.
	}
	\label{fig:compare_2d_eng_vs_t}
\end{figure*}

We also compare the global energy generation and loss rates due to neutrino emission and nuclear reactions in regions where $T < 5~{\rm GK}$ between the 1D and 2D models.
As shown in Figure~\ref{fig:compare_2d_eng_vs_t}, the total neutrino luminosities are similar in both cases, despite differences in the initial transients.
Overall, the energy loss due to neutrino emission is orders of magnitude larger than the energy generated by nuclear reactions in both the 1D and 2D cases.
The role played by nuclear reactions is similar in the 1D and 2D cases in the RGB and CBurn models: in the RGB model, nuclear heating is extremely weak in the pre-shock region, while the CBurn model is so heavily accretion-dominated that multidimensional effects are strongly suppressed.
However, for the ZACHeB and TACHeB models, the nuclear energy generation in the post-shock region is similar in magnitude but lasts longer than in the 1D models.
In the 1D simulations, the post-shock region is heated above $5~{\rm GK}$ at later times and is therefore excluded from this calculation.
In contrast, due to convective turnover blending the hot reacting material in and out of the shock region, the 2D models for these two cases exhibit longer-lasting post-shock burning and stronger pre-shock burning.

\begin{figure*}
	\centering
	\includegraphics[width=\textwidth, angle=0]{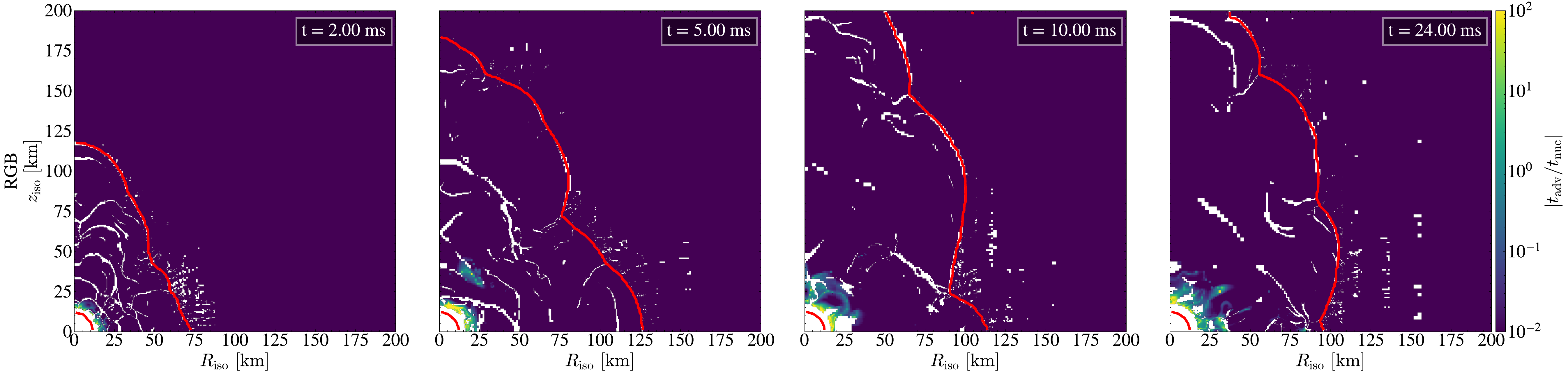}
	\includegraphics[width=\textwidth, angle=0]{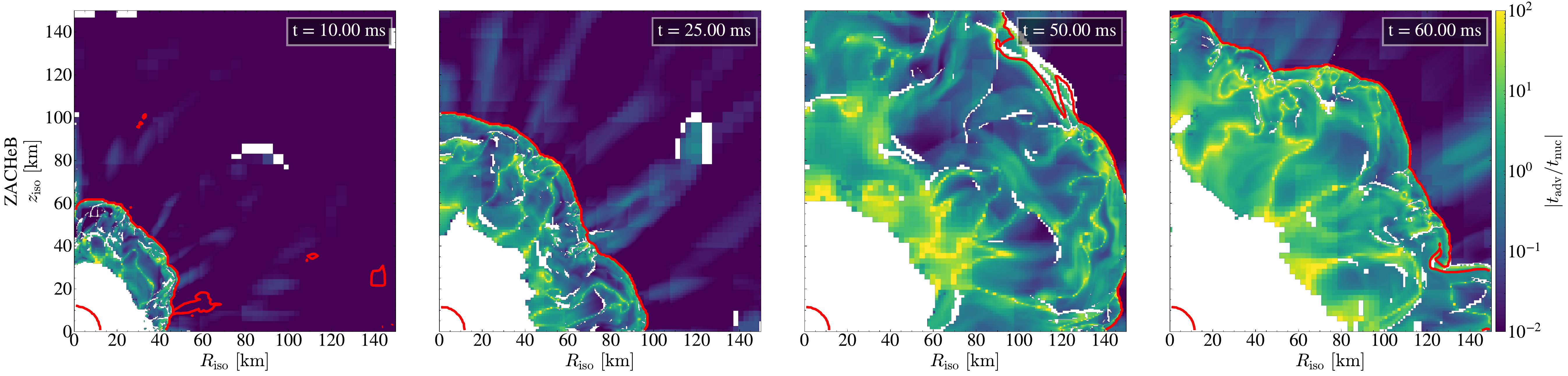}
	\includegraphics[width=\textwidth, angle=0]{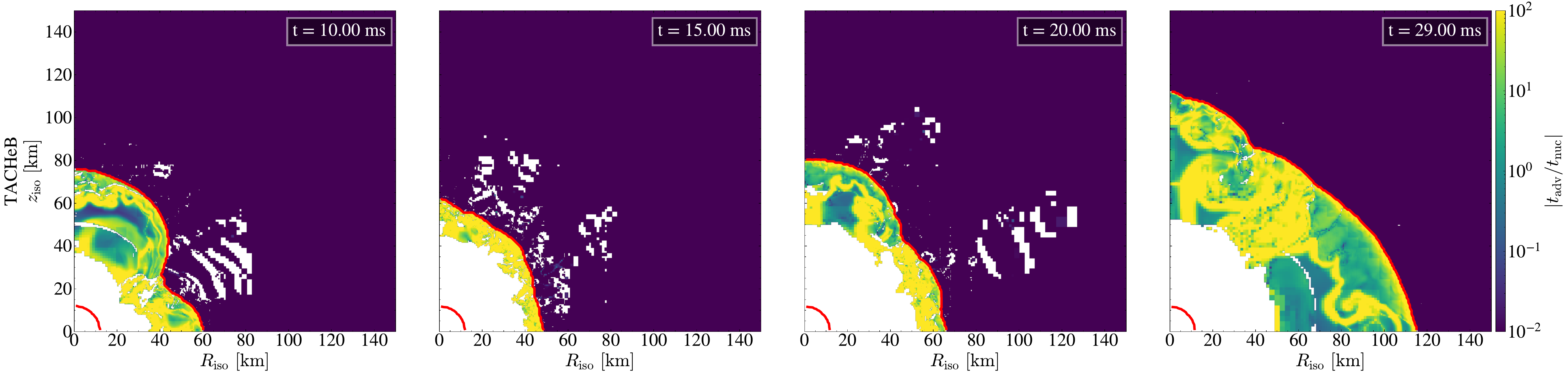}
	\includegraphics[width=\textwidth, angle=0]{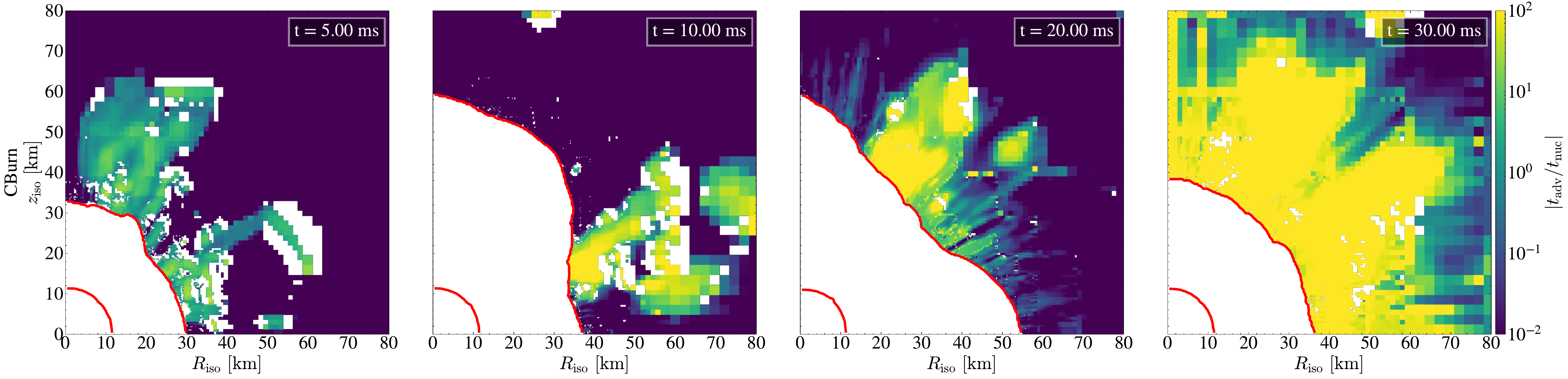}
	\caption{
		Ratio of the advection timescale to the nuclear reaction timescale, $\left| t_{\rm adv} / t_{\rm nuc}\right|$, for the different models at various times.
		Red contours mark the specific entropy $s=25~{\rm k_B per baryon}$, approximating the NS surface (lower-left region) and the shock front.
		The regions where the nuclear burning is not activated (e.g. when $T>5~{\rm GK}$ or when there is a either strong or weak shock) is marked as white.
		The snapshot times and domain sizes are identical to those in Figure~\ref{fig:2d_ent_profiles} to facilitate comparison.
	}
	\label{fig:2d_t_ratio_nuc_profiles}
\end{figure*}

While the total energy generated by nuclear burning is orders of magnitude smaller than the total neutrino energy loss, nuclear reactions can still be dynamically important locally.
To assess the dynamical importance of these nuclear reactions, we compare the advection and nuclear reaction timescales.
We locally estimate these two timescales as
\begin{equation}
    t_{\rm adv} \approx \left| \frac{\psi^2 \Delta r}{\alpha v^r} \right| , \quad
    t_{\rm nuc} \approx \left| \frac{\epsilon_{\rm int}}{\dot{\epsilon}_{\rm nuc}} \right|,
\end{equation}
where $\epsilon_{\rm int}$ is the specific internal energy.
We then compute their ratio, $\left| t_{\rm adv} / t_{\rm nuc} \right|$.
If this ratio is much larger than unity, the nuclear reaction timescale is much shorter than the advection timescale, indicating that nuclear burning cannot be dynamically ignored.
Figure~\ref{fig:2d_t_ratio_nuc_profiles} illustrates this timescale ratio at the same snapshot times as Figure~\ref{fig:2d_ent_profiles}.
As discussed above, the nuclear reaction timescale is notably short in the post-shock regions of the ZACHeB and TACHeB models.
These nuclear reactions locally deposit energy and alter the entropy gradients, thereby modulating the convective motions.
While the nuclear reaction timescale can also be very short in the CBurn model, the burning occurs primarily in the pre-shock region.
Because this region is not convectively unstable, the reactions do not drive convection and therefore have relatively little impact on the multidimensional flow dynamics.

To isolate the global impact of nuclear reactions on the accretion dynamics, we also performed short 2D simulations of the ZACHeB and TACHeB models with the nuclear network disabled (shown as dotted lines in Figure~\ref{fig:compare_2d_shock_Mdot}).
We find that the magnitude and time evolution of the angle-averaged shock radius behave very similarly to the models with nuclear burning included.
This confirms that although nuclear burning is dynamically important on a local scale by altering entropy gradients and driving convective motions, it does not significantly alter the global shock dynamics.
Because the total energy budget is vastly dominated by neutrino emission, the macroscopic behaviour of the shock is ultimately governed by the balance between the mass accretion rate and neutrino cooling.

To quantitatively evaluate the potential for successful convective dredge-up and the mass ejection of rare heavy elements, we determine whether any gravitationally unbound material has been exposed to extreme nucleosynthetic processing.
Because our Eulerian framework does not employ Lagrangian tracer particles, at each snapshot, we isolate the fluid cells that have reached temperatures of $T > 5$~GK, the threshold for NSE required to synthesize iron-peak and exotic heavy isotopes.
We then assess whether the material in these regions is gravitationally unbound.
The relativistic unbound condition is defined as 
\begin{equation}
    -h_\mathrm{th} u_t - 1 > 0,
\end{equation}
where $u_t = W(- \alpha + \beta_i v^i )$ is the covariant time component of the four-velocity, and 
\begin{equation}
    h_{\rm th} = 1 + \left(\epsilon - \epsilon_0\right) + {P}/{\rho}
\end{equation}
is the specific enthalpy including only the thermal internal energy. 
In this expression, $\epsilon$ is the specific internal energy, and $\epsilon_0$ is the corresponding zero-point energy obtained for the same rest-mass density and electron fraction but at zero temperature \citep{2020PhRvD.102l3015B, 2025ApJ...978L..38C}. 
Applying this criterion, we find that all processed material ($T > 5$~GK) remains strictly gravitationally bound in all our 2D cases throughout the simulated time.
Our finding of a completely bound inner envelope contrasts with recent multidimensional simulations by \cite{2026ApJ...997...88A}, who reported the successful ejection of processed material within 0.09~s.

Furthermore, while material may become energetically unbound at later times, this does not guarantee mass ejection if the surrounding inflow remains supersonic.
In all our 2D models, the high ram pressure of the infalling envelope creates a dynamical barrier that prevents the stalled accretion shock from expanding.
Given the sustained hypercritical accretion rates, the NS toward its maximum mass appears to be the dominant process, rendering a successful shock runoff unlikely.
Consequently, the nucleosynthetic yields remain trapped within the gravitational well and are eventually swallowed by the incipient BH.

\section{\label{sec:discussion}Discussion}


In this work, we have presented a systematic, fully coupled investigation of hypercritical accretion onto an embedded NS within massive stellar envelopes.
By combining one-dimensional and two-dimensional GRHD simulations with grey two-moment (M1) neutrino transport and an $\alpha$-chain nuclear reaction network, we tracked the thermodynamic and multidimensional evolution of the inner engine across four distinct progenitor evolutionary stages.
Our findings resolve critical ambiguities regarding the thermodynamic stability, nucleosynthetic potential, and ultimate fate of these deeply embedded systems.

\subsection{Accretion Regimes and the Validity of Steady-State Analytic Models}
Our results demonstrate that the macroscopic accretion dynamics are dictated by the progenitor's core density.
In the more compact configurations (ZACHeB, TACHeB, and CBurn), the extreme local densities ensure that the flow rapidly transitions into a quasi-steady, neutrino-cooled regime.
In this limit, our numerically determined shock radii essentially overlap with the classic idealized analytic power-law solutions \cite{1989ApJ...346..847C, 1996ApJ...460..801F}.
The minor systematic deviations we observe—namely, a slightly more compact shock—are the physically expected outcomes of our realistic hybrid equation of state and general relativistic potential, which demand steeper pressure gradients than Newtonian, purely radiation-dominated ($\gamma = 4/3$) analytic assumptions.

Conversely, the lower-density RGB models exhibit prolonged time-dependent behaviour without reaching a steady hyperaccreting state within the simulated timeframe.
This divergence implies that the steady-state assumptions ubiquitous in analytic models of embedded NSs break down for loosely bound, lower-density progenitors.

\subsection{The Global Energy Budget: Nuclear Feedback vs. Neutrino Cooling}
Our simulations clarify the delicate interplay between nuclear heating and neutrino cooling in the deeply embedded phase.
Across all simulated evolutionary stages, the total nuclear energy generation rate remains orders of magnitude smaller than the total neutrino luminosity.
However, nuclear burning proves to be highly relevant on a multidimensional, localized scale, with its dynamical impact varying drastically depending on the evolutionary state of the progenitor's core.

In earlier evolutionary stages, such as the ZACHeB phase, the primary nuclear feedback mechanism is the latent heat released during the helium burning. 
As shown by comparing our fully coupled and non-nuclear 2D simulations, this acts as a localized thermal afterburner that alters entropy gradients and significantly prolonging the duration and vigour of post-shock convective instabilities, but it leaves the macroscopic flow morphology and angle-averaged shock radius largely unchanged. 
In contrast, the advanced core composition of the TACHeB phase has richer multidimensional response.
The carbon and oxygen burning in the dense, hot post-shock region provides substantial thermal pressure and convective feedback. 
This localized energy injection is powerful enough to push the angle-averaged shock radius significantly further outward, profoundly altering the spatial extent of the accretion flow compared to models where nuclear reactions are artificially disabled.

Nevertheless, while explosive nuclear feedback can dynamically expand the accretion shock in advanced evolutionary stages, it ultimately fails to halt the infall. 
Because the global energy budget is overwhelmingly dominated by neutrino emission, the deep interior acts as an exceptionally efficient energy sink. 
Consequently, the outward mechanical work driven by even the most violent nuclear convection is entirely insufficient to alter the bulk mass accretion rate or trigger large-scale envelope unbinding, ensuring the steady accumulation of mass onto the central NS.

\subsection{Implications for T\.ZO and the Inevitability of Collapse}
The formation of a long-lived T\.ZO requires the NS to settle within the core and achieve a state of secular stability ($> 10^5$~yr).
Our fully coupled simulations suggest that such long-term stability is physically untenable in the hypercritical accretion regime.

An essential distinction between the scenario studied here and standard core-collapse supernovae (CCSNe) lies in the energy source powering the neutrino luminosity.
In a CCSN, neutrino emission is driven by the release of the binding energy of the proto-NS.
In our merger scenario, however, the neutrinos are powered entirely by the accretion of the stellar envelope.
This creates a fundamental limit on the potential for shock revival: since the energy source is the accretion flow itself, neutrino heating—even under conditions of high opacity where absorption is efficient—cannot readily exceed the ram pressure of the very material providing the power.

Consequently, the formation of a stable T\.{Z}O would require nuclear burning to successfully stall the accretion flow prior to the onset of efficient neutrino cooling.
However, our simulations demonstrate that hypercritical accretion dominates this early stage, with neutrino cooling further reducing thermal pressure support and allowing the massive stellar envelope to accrete at consistently high rates.
This effectively traps the system in a state of bounded, hypercritical accretion rather than a hydrostatic equilibrium.

To evaluate the long-term fate of the system, we extrapolate the accretion history from our long-term 1D models reported in Section~\ref{sec:1d_results}.
In the conventional $15\,M_{\odot}$ TACHeB model, the accretion rate is approximately $10^{-2}\,M_{\odot}\,\mathrm{s}^{-1}$ at $t \sim 4\,\mathrm{s}$.
Even assuming this accretion rate saturates, the NS would gain $\sim 1\,M_{\odot}$ in roughly $100\,\mathrm{s}$, inevitably crossing the Tolman-Oppenheimer-Volkoff mass limit and triggering catastrophic collapse.

Therefore, our results strongly support a dynamical paradigm: the embedded NS phase is not the birth of a stable, long-lived stellar exotic, but rather a rapid, transient precursor to catastrophic collapse.
Upon forming a BH, the subsequent accretion of the remaining massive envelope provides an ideal central engine for high-energy transients, such as ultra-long gamma-ray bursts or fast blue optical transients.

\subsection{The Failure of Convective Dredge-Up and the Fate of Heavy Nucleosynthesis}
One of the most consequential findings of our multidimensional simulations is the total suppression of mass ejection from the deep interior.
As detailed in Section~\ref{sec:results}, our Eulerian analysis reveals that all fluid elements exposed to Nuclear Statistical Equilibrium (NSE) conditions ($T > 5$~GK) remain strictly gravitationally bound.

While previous studies of the common-envelope phase (e.g., \cite{2019MNRAS.485..620K, 2026ApJ...997...88A}) have suggested that accretion feedback can successfully unbind and eject stellar material, it is crucial to distinguish those results from the deeply embedded core-merger phase presented here.
In the sparse outer envelope, the gravitational binding energy is low and neutrino cooling is weaker, allowing energy deposition to drive outflows.
In contrast, our models focus on the NS after it has reached the dense stellar core.
At these depths, the local gravitational potential is significantly deeper, and the hypercritical accretion rates trigger intense neutrino cooling that acts as an insurmountable energy sink.
Consequently, even though we observe vigorous convection, the ``ash'' from nuclear burning remains trapped within the core's deep potential well, with no viable mechanism for dredge-up or ejection.

We attribute this morphological divergence to the extreme sensitivity of the convective dredge-up mechanism to the numerical treatment of the inner boundary and neutrino microphysics.
Unlike models that terminate the inner grid exactly at the compact object surface ($r \approx 12~{\rm km}$), our domain extends to $r=8~{\rm km}$, dynamically resolving the crucial surface layer where peak neutrino-matter interactions occur.
Furthermore, self-consistently bridging to a high-density nuclear EOS, rather than employing a uniform stellar EOS, captures the fundamentally different thermodynamic and compressibility properties of matter under extreme core conditions (see, e.g., Figure~\ref{fig:1d_rho_profile}).
Within this dynamically active NS surface layer, our fully coupled M1 neutrino transport solver provides a rigorous and efficient energy sink near the neutrinosphere.
While the negative entropy gradients in the post-shock region vigorously trigger convective instabilities, the intense neutrino cooling rapidly depletes the internal energy of the rising convective updrafts.
As a result, convective bubbles lose their thermal buoyancy before they can overcome the gravitational potential of the NS.
Our dynamical results suggest that while exotic nucleosynthesis occurs within the high-temperature regions, this material is ultimately accreted rather than ejected.
This fundamentally limits the viability of hyperaccreting NSs as significant contributors of exotic isotopes to galactic chemical evolution.

\subsection{Limitations and Future Work}
While our fully coupled framework captures the dominant microphysics of the inner engine, several limitations remain.
Our 2D simulations enforce axisymmetry, which artificially constrains the turbulent cascade; full 3D simulations are ultimately required to verify the exact convective morphology.
Additionally, while we map unperturbed 1D progenitor profiles directly onto our grid, a crucial future extension will be to initialize the core-merger engine using realistic 3D profiles derived from global common envelope inspiral simulations.
Also, the use of grey two-moment neutrino transport and an approximate nuclear network introduces uncertainties in the detailed energetics and compositional evolution.
The adopted neutrino opacities are valid primarily at sufficiently high densities and temperatures, and the regions where both neutrino cooling and nuclear burning could be simultaneously important are not fully resolved in this work.
More sophisticated microphysical treatments (e.g. \cite{2024ApJS..272....9N, 2024ApJ...975..116C, 2024PhRvD.110h3028F, 2025PhRvD.111d3036C}) may refine quantitative predictions, but the qualitative trends identified in our simulations are expected to remain robust.

Furthermore, the influence of angular momentum is currently being investigated in an ongoing follow-up study, which extends this multidimensional framework to include a range of NS and progenitor rotation profiles.
This will allow for a systematic exploration of how centrifugal support and frame-dragging alter the neutrino cooling efficiency and shock dynamics.
Subsequent iterations will also incorporate general relativistic magnetohydrodynamics (GRMHD) to assess the potential for jet launch and envelope puncturing prior to the eventual collapse of the system.
Such jet-driven feedback could be a central component of the CEJSN paradigm \cite{2015MNRAS.449..288P, 2019MNRAS.484.4972S, 2021MNRAS.507.1651G, 2022RAA....22e5010S}, and determining the interplay between the detailed neutrino transport and jet-driven unbinding is crucial for predicting the resulting high-energy transients.

\begin{acknowledgments}
We thank Hiroki Nagakura for valuable discussions at the beginning stage of this project. 
P.C.-K.C. acknowledges that the initial idea for this work was conceived while he was an N3AS fellow at the University of California, Berkeley; the research was conducted and the manuscript prepared independently thereafter.

DR acknowledges support from the Sloan Foundation, from the Department of Energy, Office of Science, Division of Nuclear Physics under Awards Number DOE DE-SC0021177 and DE-SC0024388, from the National Science Foundation under Grants No. PHY-2020275, PHY-2116686, PHY-2407681, and PHY-2512802.

The simulations in this work have been performed on the third UNH supercomputer Marvin, also known as Plasma, which is supported by NSF/Major Research Instrumentation (MRI) program under grant number AGS-1919310. 
The simulations have also been performed on the Expanse cluster at San Diego Supercomputer Centre through allocation PHY230104 and PHY230129 from the Advanced Cyberinfrastructure Coordination Ecosystem: Services \& Support (ACCESS) program~\citep{10.1145/3569951.3597559}, which is supported by National Science Foundation grants \#2138259, \#2138286, \#2138307, \#2137603, and \#2138296, and on Frontera at Texas Advanced Computing Center (TACC) under allocation no.~\#PHY23001. 

We have modified \texttt{RNS}~\citep{1995ApJ...444..306S} to generate the NSs.
The Modules for Experiments in Stellar Astrophysics \citep[MESA][]{Paxton2011, Paxton2013, Paxton2015, Paxton2018, Paxton2019, Jermyn2023} (version 25.12.1) is also used to generate massive star models.
The results of this work were produced by utilising the GRMHD code \texttt{Gmunu}~\citep{2020CQGra..37n5015C, 2021MNRAS.508.2279C}, with the grey M1 neutrino transport~\citep{2023ApJS..267...38C, 2024ApJ...975..116C} and the nuclear burning~\citep{2026ApJS..284....6C} modules.
The tabulated neutrino interactions were provided by \texttt{NuLib}~\citep{2015ApJS..219...24O}.
The EoS in the non-NSE region is provided by \texttt{helmeos}~\citep{2000ApJS..126..501T}. 
The nuclear networks used in this work (i.e. the \texttt{aprox13})~\citep{1999ApJS..124..241T, 2000ApJS..129..377T} are publicly available at \url{https://cococubed.com/}.
The data of the simulations were post-processed and visualised with 
\texttt{yt}~\citep{2011ApJS..192....9T},
\texttt{NumPy}~\citep{harris2020array}, 
\texttt{pandas}~\citep{reback2020pandas, mckinney-proc-scipy-2010},
\texttt{SciPy}~\citep{2020SciPy-NMeth} and
\texttt{Matplotlib}~\citep{2007CSE.....9...90H, thomas_a_caswell_2023_7697899}.
\end{acknowledgments}

\appendix

\section{\label{sec:photon_trapping}Photon trapping}
As mentioned in section~\ref{sec:eos}, explicit radiative transfer for photons is neither necessary nor included as the simulated domain resides well inside the photon trapping radius.

The photon diffusion timescale is roughly $t_{\rm diff} \approx \kappa \rho r^2 / c$, where $\kappa \approx 0.2$~cm$^{2}$/g is the electron scattering opacity and $c$ is the speed of light.
The advection timescale is $t_{\rm adv} \approx r / |v_r|$, where $v_r$ is the radial fluid velocity.
We can evaluate this trapping condition in both the unshocked free-fall region and the post-shock envelope.

First, in the pre-shock region near our outer boundary ($r \approx 1 \times 10^9$~cm), the gas rapidly accelerates into approximate free-fall with velocities $v_r \sim 10^9$~cm/s.
Even with conservative pre-shock densities (e.g., $\rho \sim 10^1$~g/cm$^3$), the advection time is roughly $t_{\rm adv} \sim 1$~s, whereas the diffusion time scales to $t_{\rm diff} \gtrsim 10^7$~s.
While the fluid velocity is initially zero at the start of the simulation, the dynamical timescale required to establish the free-fall flow ($t_{\rm dyn} \approx \sqrt{r^3/GM} \sim \mathcal{O}(1)$~s) is many orders of magnitude shorter than the diffusion timescale.
Thus, the hypercritical accretion flow fully establishes itself long before any meaningful radiative diffusion can occur, and any photons present in the incoming stellar material are rapidly swept inward.

Second, in the post-shock region near the NS ($r \sim 10^7$~cm), the radial infall velocity is significantly reduced by the shock and convective turbulence, while the density rises sharply ($\rho \gtrsim 10^7$~g/cm$^3$).
Although the slower advection increases $t_{\rm adv}$ slightly, the massive increase in density completely paralyses photon transport, pushing the diffusion timescale to $t_{\rm diff} \gtrsim 10^{10}$~s.

Consequently, $t_{\rm diff} \gg t_{\rm adv}$ is maintained rigorously across the entire simulation domain.
Photons are strongly trapped and completely advected with the fluid.
We also consider the potential for non-linear radiative transport via the photon bubble instability, which can facilitate energy escape in radiation-dominated accretion (e.g. \cite{1996ApJ...457L..85K, 2021MNRAS.508..617Z}). 
However, given that $t_{\rm diff}$ exceeds the simulation duration by over many orders of magnitude, even a significant reduction in effective opacity due to such instabilities would be insufficient to allow radiation to decouple from the bulk inflow.
The thermodynamic contribution of this trapped radiation field is naturally captured via the photon radiation pressure ($P_{\rm rad} \propto T^4$) and internal energy components intrinsic to our hybrid equation of state.

\section{\label{sec:appendix_compact_ns}One-dimensional simulations with a $1.8~{\rm M_{\odot}}$ NS}

To explore the impact of the depth of the gravitational potential, we consider a more compact NS.
Specifically, the initial configuration is constructed with a central mass-energy density of $\epsilon_{c}/c^2 = 7.751\times10^{14}~\rm{g \cdot cm^{-3}}$.
This yields a gravitational mass of $1.80008~{\rm M_{\odot}}$ and a corresponding rest mass of $2.02701~{\rm M_{\odot}}$. 
We performed simulations using this more massive NS for the ZACHeB, TACHeB, and CBurn progenitor models.
Because the compactness of the NS introduces only secondary quantitative effects on the overall dynamics, we present these results here to preserve the clarity of the figures in the main text.

\begin{figure*}
    \centering
    \includegraphics[width=\textwidth, angle=0]{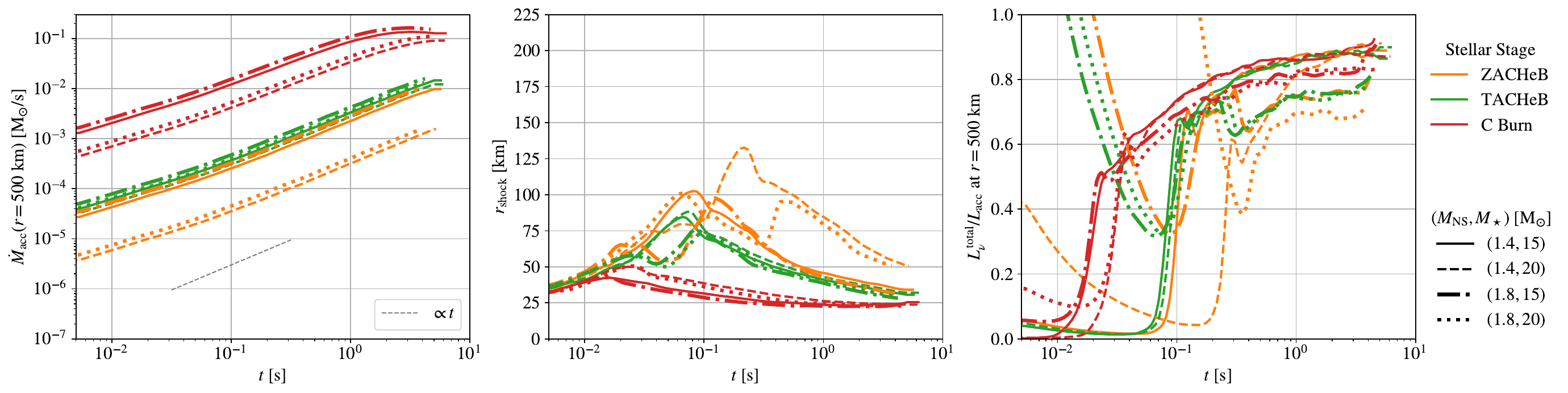}
    \caption{
        Time evolution of mass accretion rate (\emph{left}), shock radius (\emph{middle}), and the ratio of total neutrino luminosity to accretion power (\emph{right}) across all 1D models comparing different NS masses.
    }
    \label{fig:shock_Mdot_Lratio_all}
\end{figure*}

\begin{figure*}
    \centering
    \includegraphics[width=\textwidth, angle=0]{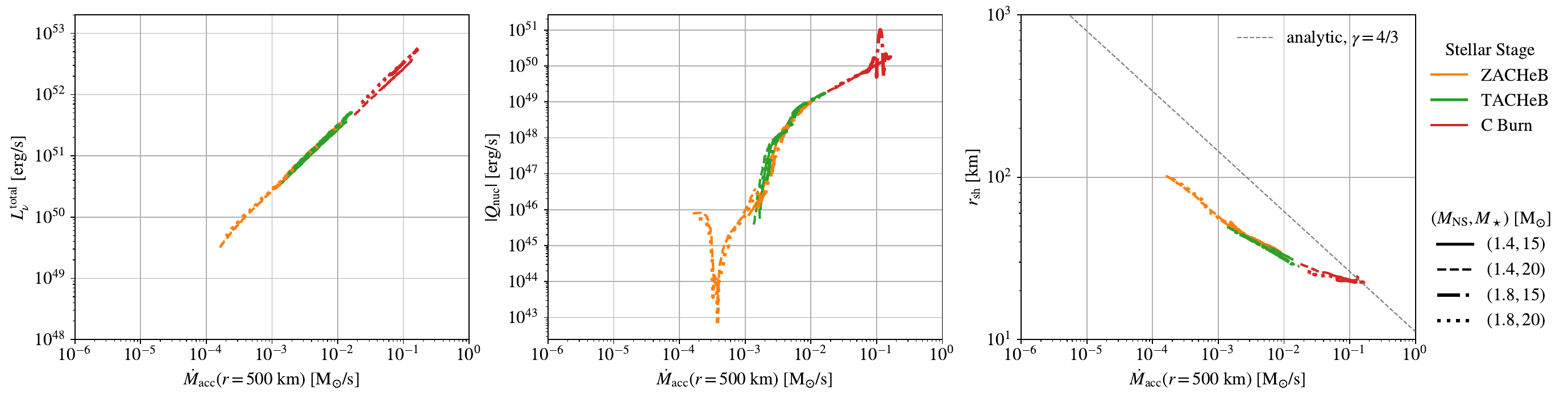}
    \caption{
        Total neutrino luminosity (\emph{left}), total nuclear energy generation rate (\emph{middle}), and shock radius (\emph{right}) as functions of the mass accretion rate for the more compact NS.
        Only late-time evolution ($t > 0.5~{\rm s}$) is shown.
        The analytic shock radii estimated by \cite{1989ApJ...346..847C, 1996ApJ...460..801F} are shown as grey lines for comparison.
    }
    \label{fig:eng_rsh_vs_mdot_all}
\end{figure*}

As shown in Figure~\ref{fig:shock_Mdot_Lratio_all}, the more compact NS induces slightly higher mass accretion rates, accompanied by moderately smaller shock radii and a slightly reduced neutrino radiation efficiency.
Nevertheless, as demonstrated in Figure~\ref{fig:eng_rsh_vs_mdot_all}, the macroscopic behaviour and the qualitative scaling with the mass accretion rate remain identical to the baseline $1.4~{\rm M_{\odot}}$ cases.

\bibliography{references}{}

\end{document}